\documentclass[traditabstract]{aa}

\usepackage{graphicx}
\usepackage{txfonts}
\usepackage{natbib}
\usepackage{aas_macros}
\usepackage{lscape}
\usepackage{longtable}

\usepackage{soul}

\usepackage[usenames,dvipsnames]{color}

\begin{document}

\title{Single stellar populations in the near-infrared\\ II. Synthesis models}
\titlerunning{Single stellar populations in the NIR - II. Synthesis models}


\author{S. Meneses-Goytia\inst{1} 
\and 
R. F. Peletier\inst{1} 
\and 
S. C. Trager\inst{1} 
\and 
A. Vazdekis\inst{2,3} 
}
\authorrunning{S. Meneses-Goytia et al.}

\institute{
Kapteyn Instituut, Rijksuniversiteit Groningen, Landleven 12, 9747AD Groningen, The Netherlands \\
\email{s.meneses-goytia@astro.rug.nl}
\and
Instituto de Astrof{\'i}sica de Canarias, via L{\'a}ctea s/n, La Laguna, Tenerife, Spain
\and
Departamento de Astrof{\'i}sica, Universidad de La Laguna, 38205 La Laguna, Tenerife, Spain
}

\date{Accepted XXXX. Received XXXX; in original form XXXX}

\abstract{
We present unresolved single stellar population synthesis models in the near-infrared (NIR) range. The extension to the NIR is important for the study of early-type galaxies, since these galaxies are predominantly old and therefore emit most of their light in this wavelength range. The models are based on a library of empirical stellar spectra, the {\it NASA infrared telescope facility (IRTF) spectral library}. Integrating these spectra along theoretical isochrones, while assuming an initial mass function (IMF), we have produced model spectra of single age-metallicity stellar populations at a resolution $R~\sim 2000$. These models can be used to fit observed spectral of globular clusters and galaxies, to derive their age distribution, chemical abundances and IMF.
The models have been tested by comparing them to observed colours of elliptical galaxies and clusters in the Magellanic Clouds. Predicted absorption line indices have been compared to published indices of other elliptical galaxies. The comparisons show that our models are well suited for studying stellar populations in unresolved galaxies. They are particularly useful for studying the old and intermediate-age stellar populations in galaxies, relatively free from contamination of young stars and extinction by dust. These models will be indispensable for the study of the upcoming data from JWST and extremely large telescopes, such as the E-ELT.
}


\maketitle


\label{firstpage}

\section{Introduction}
\hspace{0.45cm}

To understand the formation and evolution of the Universe, we analyse the light emitted by observed objects, like galaxies. This light is the spectral energy distribution (SED) and provides insights into, for example star formation histories, chemical abundances, and the distribution of mass in stars. By studying galaxies in diverse environments at different redshifts, we can understand the mechanisms that drive galaxy formation and evolution.

Galaxies beyond the Local Group are typically studied by interpreting their integrated light, rather than the light of individual stars. An approach used to study these unresolved galaxies is through evolutionary population synthesis modelling. This method \citep[e.g.][]{tinsley_1980,bruzual_and_charlot_2003,vazdekis_et_al_2010} is based on the assumption that galaxies consist of a number of single stellar populations (SSPs). Each SSP represents a single burst of star formation with a uniform initial chemical composition. With this technique it is possible to produce SEDs which can be used to derive physical properties from observations allowing determinations of star formation tracers, stellar content and evolution, chemical abundances, initial mass function (IMF) slopes, and other characteristics \citep[e.g.][]{gonzalez_et_al_1993,peletier_1993,trager_et_al_2000a,yamada_et_al_2006}. 

These studies have mainly focused on the optical wavelength range leaving the near-infrared (NIR) range almost unexplored. Observations of both stars and galaxies taken in these wavelengths are scarce and therefore, so are the stellar and SSP models. The NIR presents alternatives and opportunities that are not available in the optical, because the NIR is highly sensitive to K and M stars and is less affected by hot young stars and by dust extinction. Due to its sensitivity to cool stars, the NIR is well suited to study specific stellar populations, such as stars on the asymptotic giant branch (AGB), especially the thermally pulsating AGB (TP-AGB), and on the tip of red giant branch (RGB) stars \citep{salaris_et_al_2014}. This makes the NIR particularly attractive for studying intermediate-age galaxies ($0.5 - 2.0~\mathrm{Gyr}$), whose light is mostly due to TP-AGBs \citep{maraston_2005,marigo_et_al_2008,bruzual_et_al_2014}. 

Current models in the NIR \citep{mouhcine_and_lancon_2002,maraston_et_al_2009b} are based on available empirical \citep{pickles_1998,lancon_and_wood_2000} and theoretical libraries \citep[e.g.][]{lejeune_et_al_1997,lejeune_et_al_1998,westera_et_al_2002}. Neither type of library, however, is ideally suited for stellar population modelling. Theoretical spectra have considerable problems in reproducing molecular bands \citep[e.g.][]{martins_and_coelho_2007}, and since these are dominant in many cool stars, they cannot be used at present to make accurate predictions for absorption line indices in the NIR. The empirical library of \citet{lancon_and_wood_2000} has a spectral resolution ($R~\sim 1000$), but is limited to cool stars; hence \citet{mouhcine_and_lancon_2002} also include theoretical stars in their stellar population models. The models from \citet{maraston_et_al_2009b}, on the other hand, are entirely based on a theoretical stellar spectral library. Given these limitations, the observations made by \citet{rayner_et_al_2009} and \citet{cushing_et_al_2005}, compiled in the {\it IRTF spectral library}, have allowed us the opportunity to improve SSP models at $R \sim 2000$, containing a larger sample of cool and late-type stars than its predecessors. \citet{conroy_and_gunn_2009, roeck_et_al_2015} used some spectra of this library, showing the advantages of using this library in the NIR.

Our aim is to offer an improved tool for stellar population studies in the NIR $J$, $H$, and $K$ bands ($0.94 - 2.41~\mu m$). This will be particularly valuable for future science conducted with the observations of telescopes such as E-ELT and JWST, which will focus on the NIR. Hence, we present the following series of papers. In the first paper \citep[][hereafter Paper I]{paper_I} we characterised one of the model ingredients, the {\it IRTF spectral library}, by determining the stellar parameters of the stars and the resolution of the library, as well as the reliability of the flux calibration. In Paper II, the present work, we introduce our modelling approach and its predictions. The third paper \citep[][hereafter Paper III]{paper_III} will show the comparisons of our models with observations of early-type galaxies from full-spectrum fitting and line-strength index fitting approaches.

In this paper, we present our SSP models in the NIR range. In Section 2 we describe the construction of our models, which follow a similar approach to \citet{vazdekis_et_al_2010}, and also describe the ingredients. The model predictions and a discussion are given in Section 3. Finally in Section 4, we present a summary and final remarks.


\section{Single stellar population synthesis models}
\label{models_synthesis}
\begin{figure}
	\centering
	\includegraphics[width=\columnwidth]{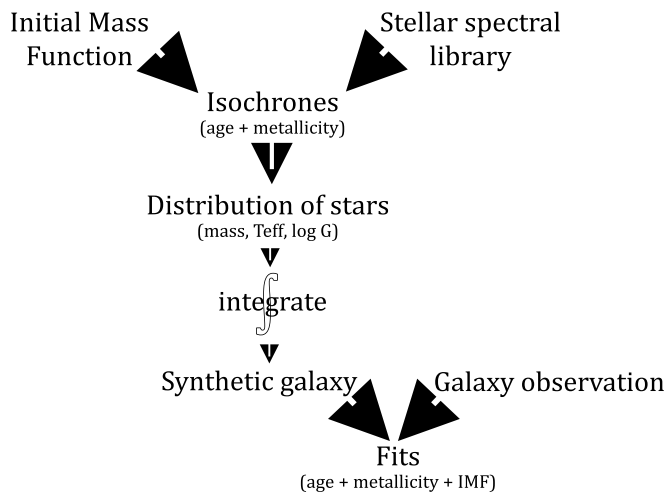}
	\caption{Basic outline for SSP modelling, after \citet{tinsley_1980}.}
	\label{ssp_model}
\end{figure}

\hspace{0.45cm}
Stellar population synthesis is a powerful technique for studying galaxy evolution, allowing us to determine galaxy ages and chemical abundances. Using a given set of isochrones at a certain age and metallicity and an assumed Initial Mass Function [IMF, $\Phi(m)$], we find for every point in the isochrone with a given effective temperature ($T_\mathrm{eff}$), gravity ($\log g$) and metallicity ($[Z/Z_{\odot}]$), a spectrum of a star by interpolating the spectra of the stellar library. With this we obtain individual stellar spectrum of a distribution of stars, which we integrate weighting each spectrum by its luminosity in a chosen wavelength and the number of stars given by the IMF in that mass bin. We then obtain a synthetic spectrum (SED) of a galaxy with a particular set of parameters (i.e. age and metallicity). A general scheme of the steps taken in SSP modelling are shown in Figure\,\ref{ssp_model}, based on \citet{tinsley_1980}.

SEDs provide a versatile tool for analysing observables, because they can be compared to observed galaxy spectra (when convolved to the adequate resolution) or used to obtain specific information, such as integrated colours or line-strength indices. This kind of constructing approach has been previously employed by \citet{vazdekis_et_al_2010,vazdekis_et_al_2003} in the optical range using the MILES \citep[3540 - 7410~\AA,][]{sanchez-blazquez_et_al_2006a} and CaT \citep[8349 - 8952~\AA,][]{cenarro_et_al_2001c} empirical stellar libraries.

The SSP SED, $S_{\lambda}$(t,[Z/Z$_{\odot}])$ is calculated with the following prescription:

\
\begin{small}
\begin{equation}
	S_{\lambda}(t,[Z/Z_{\odot}])=\int_{m_1}^{m_t}\!S_{\lambda}(m,t,[Z/Z_{\odot}])N(m,t)F_{H}(m,t,[Z/Z_{\odot}])\,dm\,
	\label{eq1}
\end{equation}
\end{small}
\

\begin{equation}
	[Z/Z_{\odot}]=\log(Z/Z_{\odot})
	\label{eq2}
\end{equation}	
\
	
\begin{equation}
	N(m,t)={\Phi(m)}{\Delta}m
	\label{eq3}
\end{equation}
\

\begin{figure}[!t]
	\centering
	\includegraphics[width=\columnwidth]{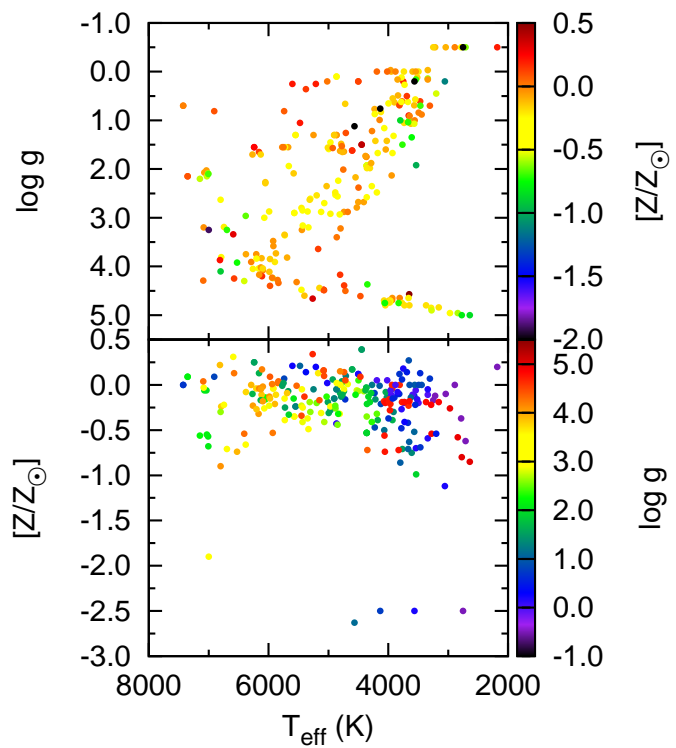}
	\caption{Parameter coverage of the {\it IRTF spectral library} (values from Paper I).}
	\label{irtf_classes}
\end{figure}

\noindent Here $S_{\lambda}(m,t,[Z/Z_{\odot}])$ is the empirical spectrum which corresponds to a star of mass $\it{m}$ and metallicity $[Z/Z_{\odot}]$  \citep[where $Z_{\odot} = 0.019$ from][]{grevesse_et_al_2007} alive at the age $t$ assumed for the stellar population $t$, $N(m,t)$ is the number of stars of mass $m$ at age $t$, which depends on the adopted IMF for the galaxy, $\it{m_1}$ and $\it{m_t}$ are the stars with the smallest and largest stellar masses, respectively, which are alive in the SSP (the upper mass limit depends on the age of the stellar population), and $F_{H}(m, t , [Z/Z_{\odot}])$ is the flux of the star in the $H$ band. Before integrating the spectra of the stars, each requested stellar spectrum is flux normalised by convolving its flux with the filter response curve of the $H$ band from the photometric system of \citet{bessel_and_brett_1988}. In this way, we ensure that each spectrum has the correct luminosity as given by the isochrone.

To obtain the empirical spectrum of each star (point) in the isochrone, we followed a similar interpolation scheme as that used in Paper I (see Section 3.1). The interpolator uses our set of input stellar spectra to interpolate (or extrapolate) to the desired values of stellar parameters. In order to improve the interpolation in regions with sparse parameter coverage (e.g. cool bright giants), a subset of interpolated stellar spectra is used as secondary input sources. After the first iteration, we selected the synthetic stellar spectra for which the $T_\mathrm{eff}$ agrees with the input value within $1\sigma$ ($\sim 200~\mathrm{K}$). These interpolated stars are then added to input library after which a second round of interpolation is performed on those stars with $\Delta{T_\mathrm{eff}} \ge 1\sigma$. Subsequently, the $(J-K)$ colour of this subset is compared to the expected colour determined from the relevant isochrone. All stars  with $\Delta{(J-K)} \le 1\sigma$ ($\sim 0.1$ mag) are classified as useful stars and are added to the input library, after which a final round of interpolation is conducted for the remaining stars. This approach was particularly useful for phases of the isochrone where the {\it IRTF spectral library} presents a deficiency of parameter cover like cool bright giants.

\begin{figure}[!t]
	\centering
	\includegraphics[width=\columnwidth]{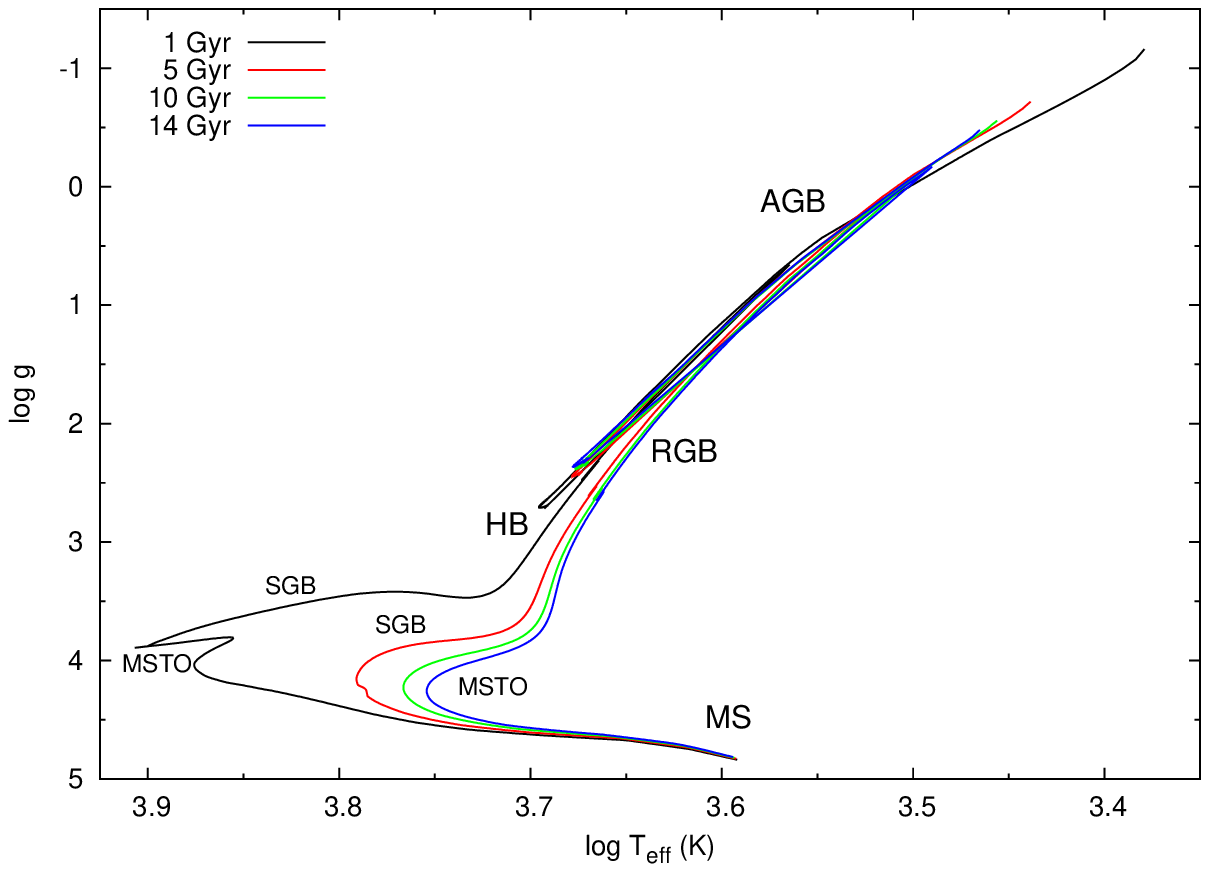}
	\includegraphics[width=\columnwidth]{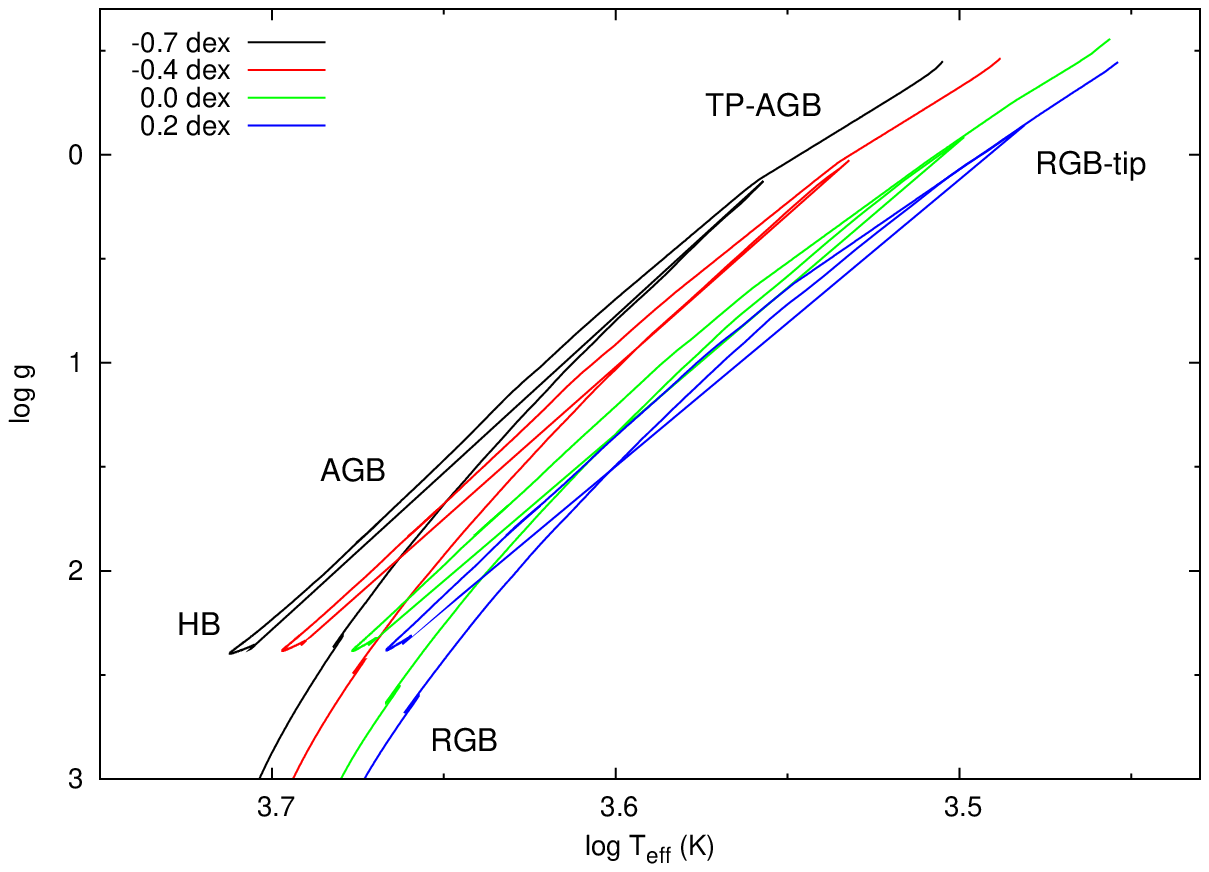}
	\caption{Location of the different evolutionary stages in the BaSTI isochrones.}
	\label{isochrones_sequences}
\end{figure} 

\begin{figure}[!ht]
	\centering
	\includegraphics[width=\columnwidth]{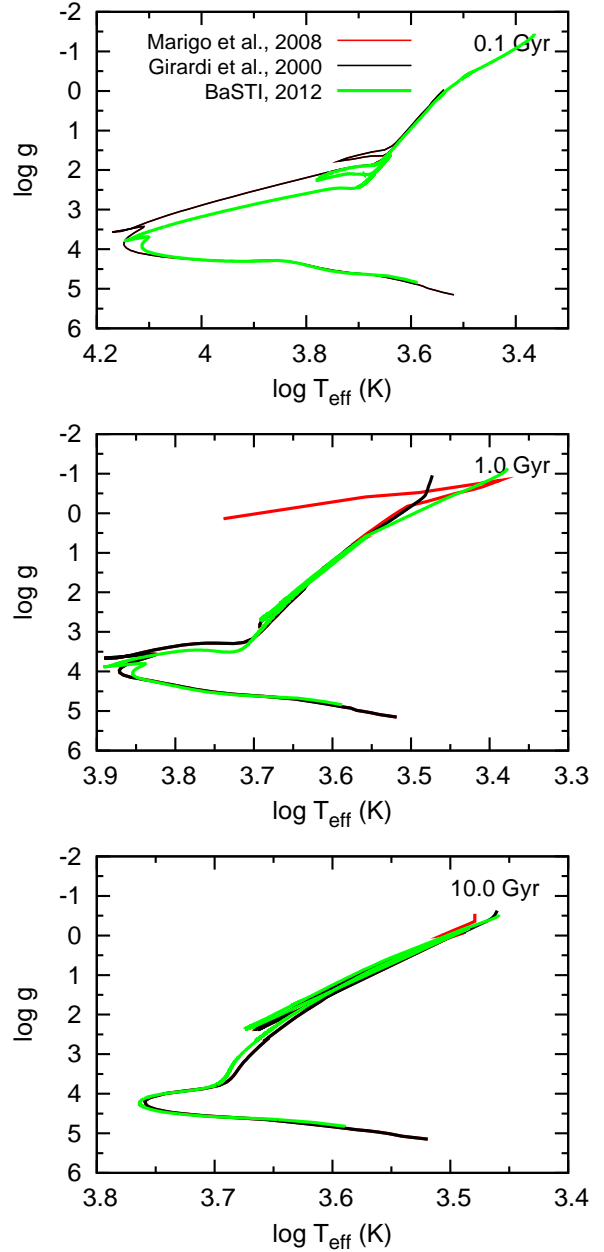}
	\caption{Comparison of different isochrone sets used, at solar metallicity and different ages.}
	\label{compare_isochrones_ages}
\end{figure}

In this work, we choose an empirical stellar library in the NIR (Section \,\ref{sec_irtf}), a single IMF (Section \,\ref{sec_imf}), and three different sets of isochrones (Section \,\ref{sec_iso}) which allow us to produce three sets of SSP models\footnote{Models$'$ SEDs, integrated colours and line-strength indices available on-line at \url{smg.astro-research.net}}. In Table\,\ref{models_info}, we present these ingredients. 


\subsection{The IRTF spectral library}
\label{sec_irtf}
\hspace{0.45cm}
A stellar spectral library is a crucial ingredient of SSP models since it provides the behaviour of spectra of individual stars as function of effective temperature, gravity and composition. Depending on the wavelength coverage of the libraries, we can analyse different stellar phases. Here we chose a stellar library with a wavelength range covering the $J$, $H$ and $K$ bands of the Near-infrared. In these regions, AGB stars dominate the light of intermediate age populations and RGB the light of old stellar populations \citep{frogel_et_al_1988,maraston_2005,conroy_and_gunn_2010}.

\begin{figure*}[!ht]
	\centering
	\includegraphics[width=\textwidth]{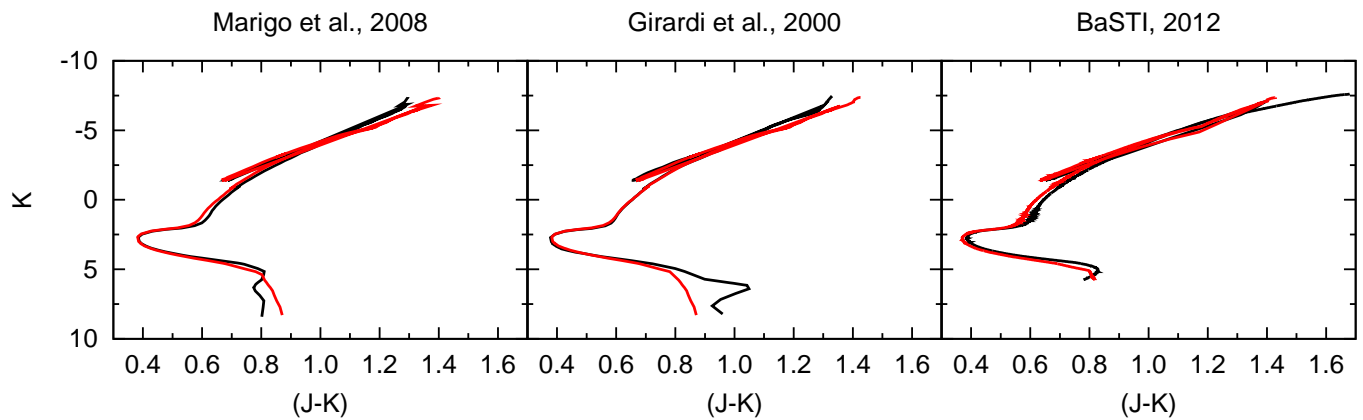}
	\caption{Comparison of colour-magnitude changes of the isochrones due to our parameter-colour empirical transformations, in a $10~\mathrm{Gyr}$ isochrone at solar metallicity. The black line represents the original isochrones and the red line the transformed isochrone.}
	\label{compare_isochrones}
\end{figure*} 

\begin{figure*}
	\centering
	\includegraphics[width=\textwidth]{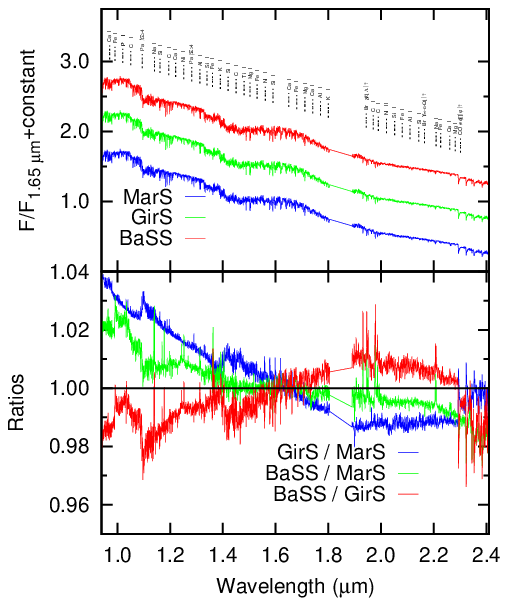}
	\caption{SED of our three SSP models at solar metallicity and $10~\mathrm{Gyr}$ (upper panel) and ratios when comparing to each other (lower panel). Interesting spectral features \citep[from][]{rayner_et_al_2009} are marked in the upper panel.}
	\label{full_seds}
\end{figure*}

We use the empirical stellar spectra library observed with the medium-resolution spectrograph SpeX at the NASA Infrared Telescope Facility on Mauna Kea \citep{rayner_et_al_2009, cushing_et_al_2005}. This library is known as the {\it IRTF spectral library} and is a compilation of stellar spectra that fall into two ranges of the IR, from which we only focus on the spectral region for the $J$, $H$ and $K$ bands ($0.94 - 2.41~\mu m$). These spectra were observed at resolving power of $2000$ ($R=\lambda/\Delta\lambda$) and were not continuum-normalised allowing for the retention of the strong molecular absorption features of cool stars. Keeping this shape also allowed absolute flux calibration by scaling the spectra to published Two Micron All Sky Survey (2MASS) photometry. For details on the observations, data reduction and calibrations, we refer the reader to  \citet{cushing_et_al_2005} and  \citep{rayner_et_al_2009}. The integrated colours obtained from the spectra are consistent with those obtained by 2MASS (see Paper I for more information). An atmospheric absorption band between $1.81$ and $1.89~\mu$m causes a loss of flux in the $H$ band. This loss is accounted for below in Section \,\ref{sec_iso}.

The 210 stars that constitute this library include late-type stars, AGB, carbon and S stars. Spectral types range from F to T and luminosity classes from I to V, as shown in Figure \,\ref{irtf_classes}. 

In Paper I, we determined the stellar parameters of the stars. Most of these stars have metallicities close to solar but stars with metallicities down to $-2.6~\mathrm{dex}$ are also included. However, we adopt a lower limit of $[Z/Z_{\odot}]=-0.70~\mathrm{dex}$ in the generation of SEDs since lower metallicities are not fully sampled (Figure 11 of Paper I). Figure\,\ref{irtf_classes} presents these parameters and demonstrates a problem of most empirical libraries, including the {\it IRTF spectral library}: a deficiency of very cool, low-mass stars and cool bright giants.\citep[e.g.][]{vazdekis_et_al_2010,conroy_and_van_dokkum_2012}. Therefore, it is important to bear in mind that the models for metallicities $-0.70~\mathrm{dex}$ and $0.20~\mathrm{dex}$ are made using considerable extrapolations, which means that they may be less accurate than the other models. At the low metallicity end, the isochrones are not easily populated at the main-sequence (MS) and MS turn-off (MSTO) however, these warm stars are easier to extrapolate from the stellar library. At the high metallicity end, where cool bright giants number is low, the interpolation scheme previously described is crucial to overcome this limitation.


\subsection{Initial mass function}
\label{sec_imf}
\hspace{0.45cm}   
The IMF describes the mass distribution of stars formed in one single burst. In this work, we adopt a power-law IMF described by \citet{salpeter_1955} using a mass range of $0.15 - 100~M_{\odot}$.

In this paper we neither defend nor argue against the universality of the IMF. Nonetheless, in future work, we will address the effect of different IMFs on the SEDs, integrated colours and line strength indices.

   
\subsection{Isochrones}
\label{sec_iso}
\hspace{0.45cm} 
An isochrone is the locus in the Hertzsprung-Russell diagram of all stars with the same initial chemical composition and with same age. The shape of the isochrone depends on the adopted prescriptions for stellar evolution. As previously mentioned, the NIR is very sensitive to stellar populations that contain TP-AGB and RGB stars, therefore it is important to analyse the impact that the different prescriptions that are used to compute stellar evolution have on SSP models by using three different sets of isochrones from two different groups. As a reference, in Figure \,\ref{isochrones_sequences} we show the \citet[][hereafter BaSTI]{pietrinferni_et_al_2004} isochrones at solar metallicity and different ages to be used as a reference for locating the different evolutionary stages in the isochrones.

\begin{table}
	\begin{center}
	\caption[]{Single Stellar Population model properties and parameter coverage}
         \label{models_info}
	\resizebox{0.5\textwidth}{!}{
	
          \begin{tabular}{@{}lccc}
          \hline
		{\bf Property}			& {\bf Characteristics}						& Notes \\
	 	\hline
		\hline
		Stellar library			& {\it IRTF spectral library}	 					& \\
               	Wavelength range		& 0.93 - 2.41~$\mu m$						& $J$, $H$ and $K$ bands \\
		Resolution			& 5.9 \AA~at 1.22~$\mu m$					& See Paper I \\
							& 7.6 \AA~at 1.62~$\mu m$					& \\
							& 9.3 \AA~at 2.02~$\mu m$					& \\
							& 9.7 \AA~at 2.27~$\mu m$					& \\
		Wavelength frame		& vacuum									&  \\
							& 										&  \\
		IMF					& \citet{salpeter_1955}						& S \\
		Stellar mass range		& 0.15 - 100.0~M$_{\odot}$					& \\
          						& 										&  \\
		Metallicity range		& $-$0.70~$\mathrm{dex}~{\le}~[Z/Z_{\odot}]~{\le}~0.20~\mathrm{dex}$ 		& See text for discussion \\
							& 										& of the models at limits \\
							& 										&  \\
		Age range			& 1 - 14~$\mathrm{Gyr}$ 						& \\
							& 										&  \\
            	Isochrones			&  \citet{marigo_et_al_2008} 					& Mar \\  
							&  \citet{girardi_et_al_2000} 					& Gir \\ 
							&  \citet{pietrinferni_et_al_2004}				& BaS \\
            	\hline
         \end{tabular}
         }
         \end{center}
\end{table}         

For one set of models, we use isochrones developed by \citet[][hereafter G00]{girardi_et_al_2000}. These evolutionary tracks were computed with updated opacities and equation of state, including convective overshooting. The evolutionary phases go from the zero age main sequence to the Thermally Pulsating Asymptotic Giant Branch (TP-AGB) regime or carbon ignition. 

\begin{figure*}[!ht]
	\centering
	\includegraphics[width=\textwidth]{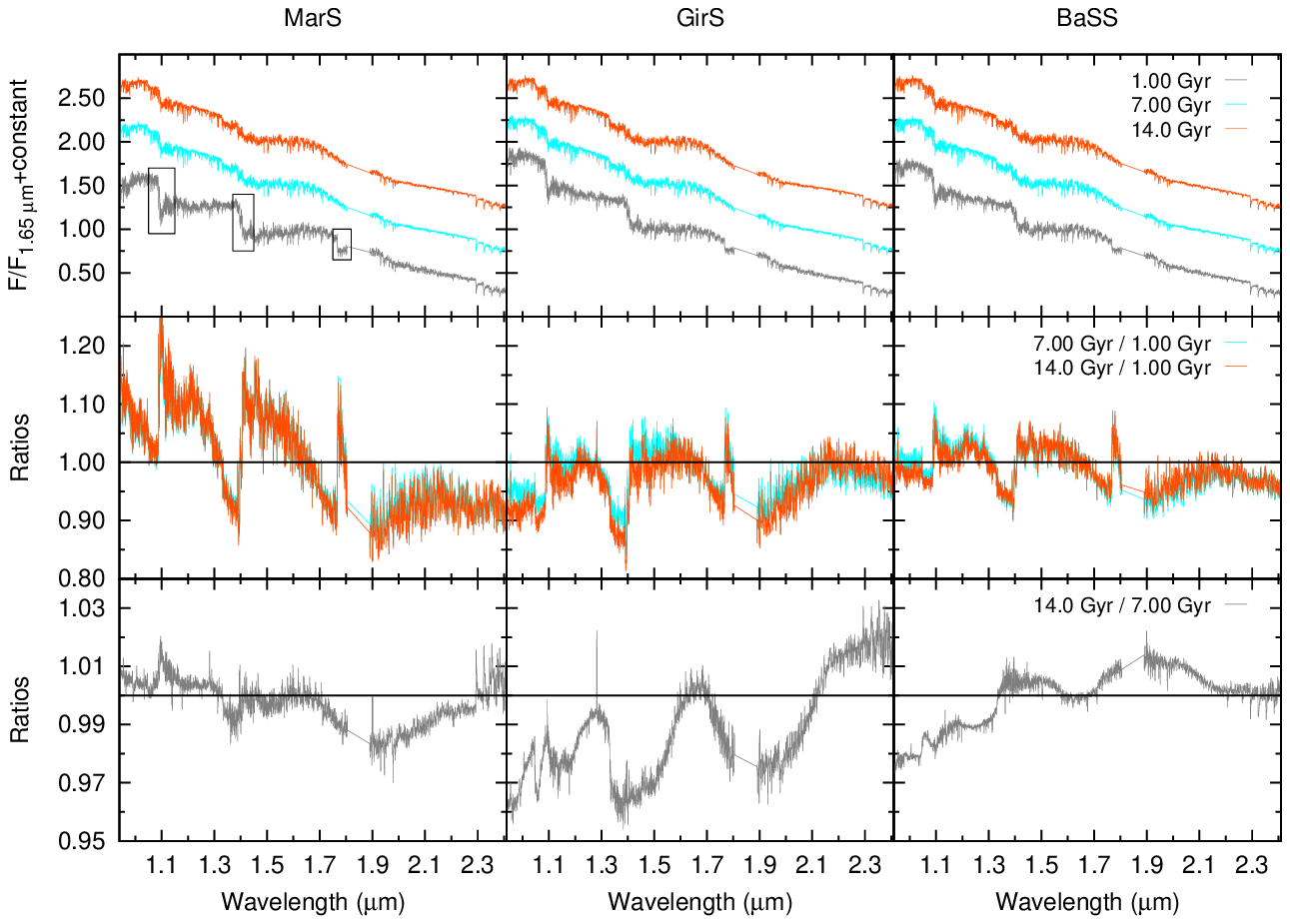}
	\caption{SEDs comparisons of our SSP models at solar metallicity and different ages (upper panel) and their ratios when comparing the models to each other (lower panel). The distinctive features of C-stars are indicated by boxes in the $1~\mathrm{Gyr}$ SED.}
	\label{seds_metal_ages}
\end{figure*}

\begin{figure*}[!ht]
	\centering
	\includegraphics[width=\textwidth]{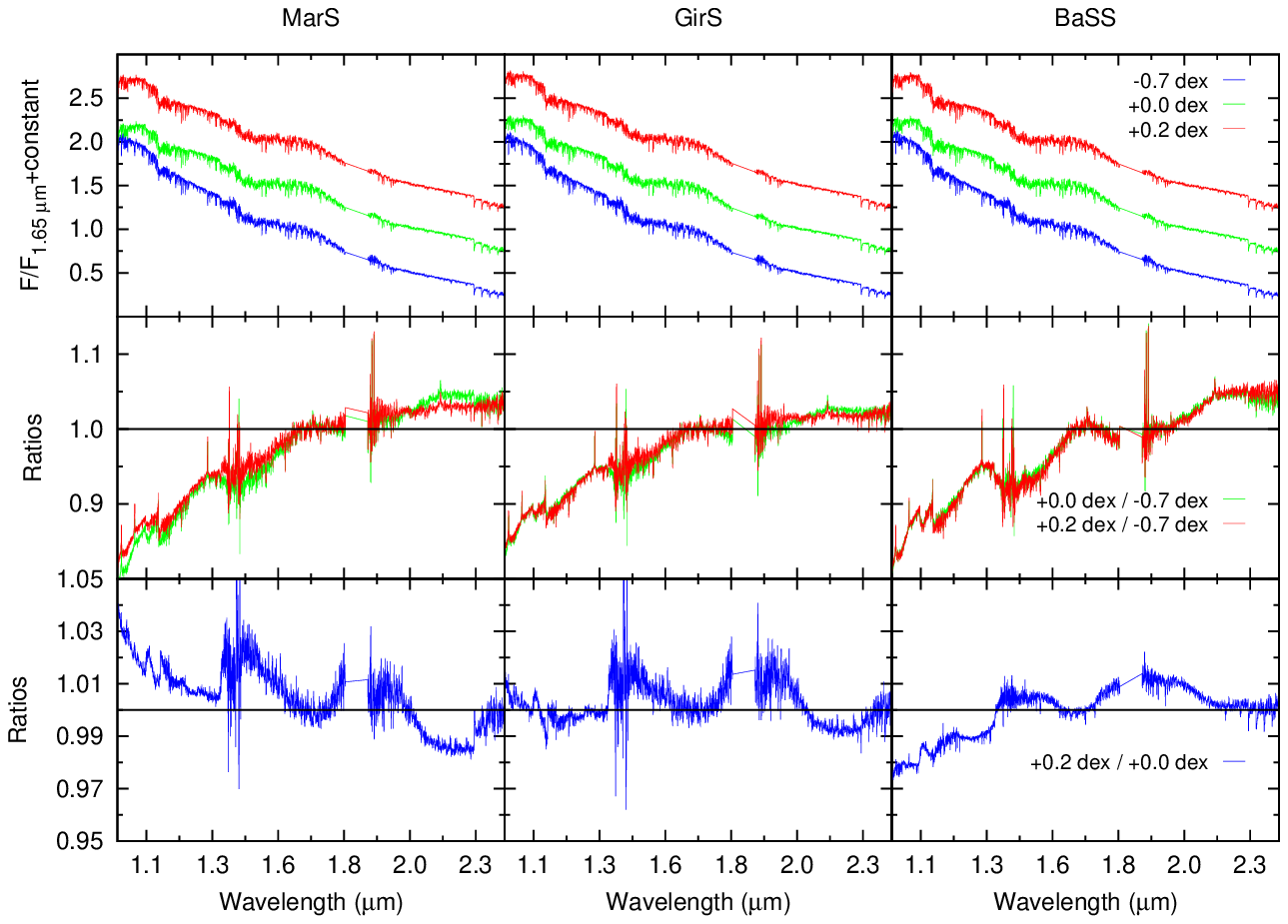}
	\caption{SEDs comparisons of our SSP models at $7.0~\mathrm{Gyr}$ and different metallicities (top panels) and their ratios when comparing the models to each other (lower panels).}
	\label{seds_age_metals}
\end{figure*}

Another set of models use the isochrones developed by \citet[][hereafter M08]{marigo_et_al_2008}. These isochrones improved the physical processes used by their predecessors (G00), for TP-AGBs by using variable molecular opacities instead of the scaled-solar tables for complete AGB models. Their calibration allows the isochrones to be fairly reliable for the TP-AGB phase and have reasonable lifetimes at metallicities of the Magellanic Cloud (MC) clusters. 

The third set of isochrones used are the AGB-extended isochrones of BaSTI. These cover the full thermally pulsating phase, using the synthetic AGB technique at the beginning of the TP phase, where the full evolutionary models stop. The TP-AGB phase is then followed by increasing the CO core mass and luminosity. Mass loss of the envelope is included and when this reaches a negligible mass, the synthetic evolution stops. After this point,  the tracks evolve at constant luminosity towards their white dwarf cooling sequences. We note that the BaSTi isochrones as published do not include stars with $M<0.5~M_\odot$. The lack of low-mass dwarfs in the BaSTI-based models may cause some discrepancies with other models, in for example, IMF-sensitive spectral features. The contribution of these dwarfs to the NIR light is $\sim 20~\%$ \citep[see also, e.g.][]{frogel_1988}.

In Figure \,\ref{compare_isochrones_ages}, we compare the three sets of isochrones at solar metallicity and three ages. It is noteworthy that in all three isochrone sets, the temperature of the MSTO exceeds $7500~\mathrm{K}$, the temperature of the hottest stars in the {\it IRTF spectral library}, for ages younger than $1.0~\mathrm{Gyr}$. We therefore only model SSPs with ages from $1.0$ to $14.0~\mathrm{Gyr}$. When comparing the different isochrone sets at $1.0~\mathrm{Gyr}$ (middle panel), we notice the effects of the different prescriptions used by each group. For instance, the MS for M08 and G00 is virtually identical, but for BaSTI the onset of the MS is at higher temperature because it starts at higher masses. Furthermore, MSTO for BaSTI takes place at lower temperatures and lower masses than G00 and M08. At and after the MSTO, and up to the end of the subgiant branch, the three sets behave similarly. However, in the region between the RGB and the AGB, we can see the different prescriptions for the TP-AGBs: G00 and BaSTI share a similar extension to the red. Additionally, the extension to higher temperatures after the AGB termination in M08 is the onset of the white dwarf cooling sequence. Moreover, the differences when reaching older ages like $10.0~\mathrm{Gyr}$ (lower panel) are minor and only noticeable in the RGB and AGB region, while for M08, the TP-AGB phase is warmer than G00 and BaSTI but these are the populations that most affect the NIR.

An additional characteristic of all these isochrones is that the authors provide their own results for the transformation from the theoretical to observed planes. The three sets of isochrones obtain their magnitudes and colours from convolving spectra from empirical and/or theoretical stellar libraries with the response curve of several broad-band filters. By using different libraries or versions of the libraries, the isochrones have different colour-magnitude diagrams, as seen in Figure\,\ref{compare_isochrones}. Because of these differences, we use our own transformations to obtain homogeneous empirical magnitudes and to compare the results of these different sets without concern for the possible bias of the colour transformations on the results. We follow the colour-temperature relations for dwarfs and giants of \citet{alonso_et_al_1996,alonso_et_al_1999} and then apply the metal-dependent bolometric corrections of \citet{alonso_et_al_1995,alonso_et_al_1999}. We adopt $BC_{\odot}= -0.12$ and a solar bolometric magnitude of $4.70$. With this, we calculate magnitudes and fluxes in the $JHK$ photometric system of \citet{bessel_and_brett_1988}. Figure\,\ref{compare_isochrones} shows, in three different panels, a comparison of the original isochrones at $10~\mathrm{Gyr}$ and solar metallicity and those with our colour-temperature empirical transformation. The main changes in magnitudes and colours are found in the lower, un-evolved main sequence and the AGB region. 

Since in this work we focus on the NIR range, we normalise the flux in the $H$ band for the stars when integrating the stellar population. As previously mentioned in Section \,\ref{sec_irtf}, the {\it IRTF spectral library} is missing flux in the $H$ band due to an atmospheric absorption feature between $1.81$ and $1.89~\mu m$. To quantify this loss, we calculate the flux in this band in the \citet{pickles_1998} stellar spectral flux library, with and without the flux loss in the atmospheric absorption band. That gives us a loss factor independent of stellar type and  luminosity class of $\sim 0.02 \%$ in average. The loss is taken into account when we weight each interpolated stellar spectrum with its corresponding flux in $H$ band. Although this loss is negligible, we include this correction since we choose the $H$ band as the anchor of our models.

  
\begin{figure*}
	\centering
   	\includegraphics[width=\textwidth]{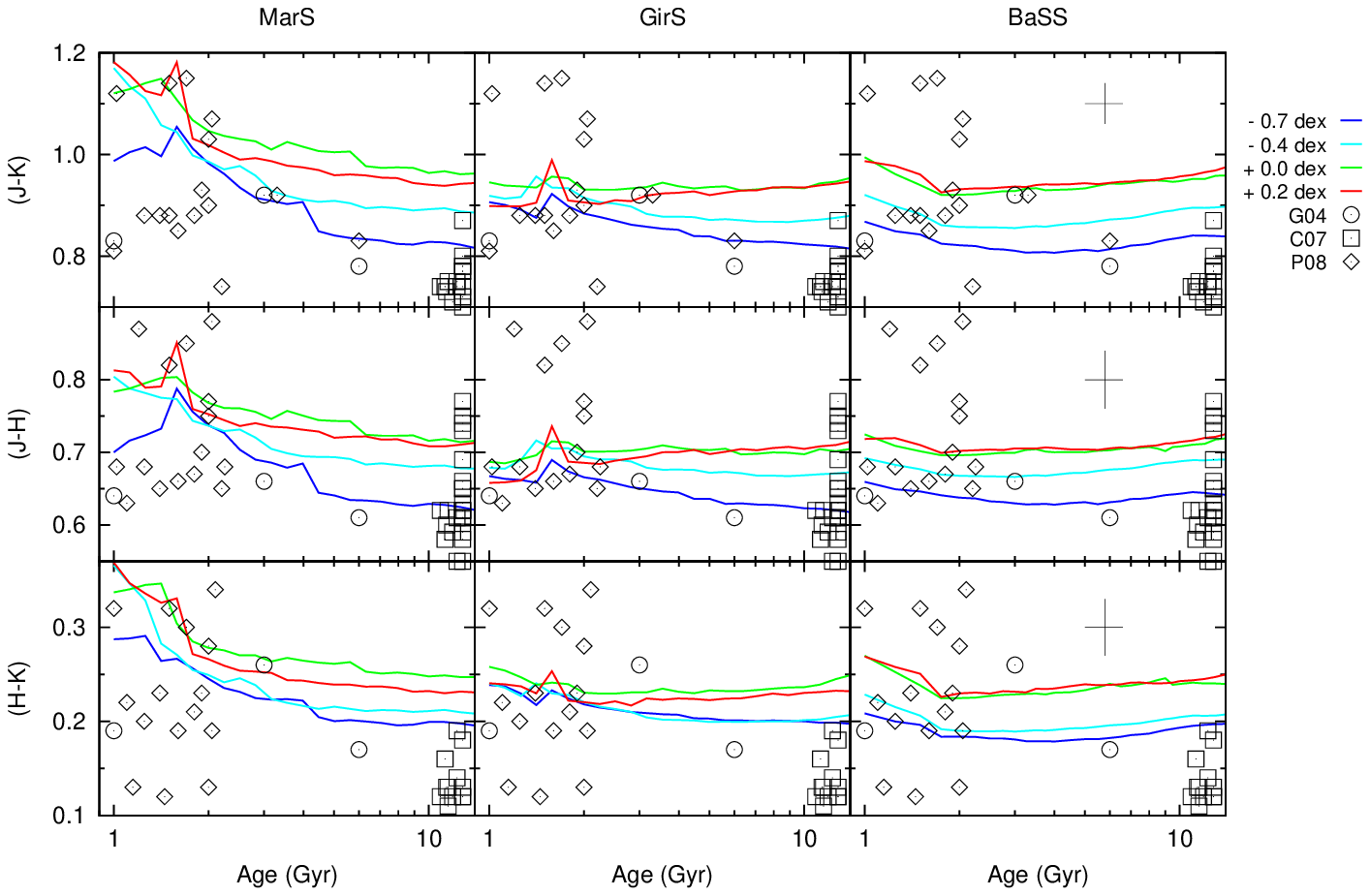}
	\caption{Comparison of integrated colours as a function of age and metallicity, from models using different sets of isochrones with Magellanic Clouds cluster observations of \citet[][G04]{gonzalez-lopezlira_et_al_2004} and \citet[][P08]{pessev_et_al_2008}, and Milky Way clusters from \citet[][C07]{cohen_et_al_2007}. These data have estimated metallicities of $\sim-0.9~\mathrm{dex}$ or higher. The error bars represent the typical uncertainties of the cluster colours and inferred ages.}
	\label{color_age}
\end{figure*}

\section{Model predictions and discussion}
\label{results}
\hspace{0.45cm}
In this section we present the results of the Single Stellar Population synthesis models created using the method described in Section \,\ref{models_synthesis}. The parameter coverage of the {\it IRTF spectral library} allows us to generate SEDs from $0.93$ to $2.41~\mu m$  (covering the $J$, $H$ and $K$ bands) for ages between $1.0$ and $14.0~\mathrm{Gyr}$ and $[Z/Z_{\odot}]$ from $-0.70$ to $0.20~\mathrm{dex}$. With a similar procedure as that used for the spectral resolution of the {\it IRTF spectral library} (Paper I), we calculate the resolution of our models. Table \,\ref{models_info} compiles the properties and parameter coverage of our models. We use the following nomenclature throughout the paper to describe our models, e.g. MarS models use the M08 isochrones (Mar) and the Salpeter (S) IMF.

It is worth pointing out the treatment we followed when C-stars are present in the population. M08 isochrones provide the C/O ratio which is a flag for carbon stars are present in young populations ($\la1.58\,\mathrm{Gyr}$). For these young MarS models, we let the interpolator use the C-stars present in the {\it IRTF spectral library} when the C/O ratio $\ge 1$. For older populations, we did not use carbon stars. On the other hand, for the GirS and BaSS models, since the isochrones do not provide the C/O ratio, we let the interpolation freely use the C-stars only for populations younger than $1.6~\mathrm{Gyr}$.

We now describe our resulting models: SEDs, integrated NIR colours and a pair of line strength indices in the $K$ band. These properties allow us to calibrate our models and establish their robustness when comparing them with well-defined observations of MC clusters. 

 
\subsection{Spectral energy distributions}
\label{results_seds}
\hspace{0.45cm}      
In Figure\,\ref{full_seds} we present SEDs of the SSP models at solar metallicity and $10.0~\mathrm{Gyr}$, using the three sets of isochrones described in Section \,\ref{sec_iso}. The main spectral features are labelled. All models are qualitatively similar, although the residuals (bottom panel) show in depth the differences and similarities between models. The GirS model presents shallower CO absorption features (between $1.60 - 1.75~\mu m$ and after $2.29~\mu m$) when compared to MarS. This is due to the stronger relative contribution of TP-AGB stars in the MarS model. Even though the BaSS model has a similar treatment of the TP-AGB phase as GirS, BaSS extends to even cooler stars and therefore has deeper CO features than both the GirS and MarS models. The BaSS models have an overall temperature difference by being cooler than the MarS and GirS models. The ratios shown in this figure (lower panel) indicate differences smaller than $2 - 3\%$, which implies a limited impact of the varying isochrones (for the old populations). 

Figure\,\ref{seds_metal_ages} presents solar metallicity models at different ages ($1$, $7$ and $14~\mathrm{Gyr}$). The upper panels show the SEDs and the lower ones present the ratios when comparing different ages. All three models at $1~\mathrm{Gyr}$ show the presence of C-stars that have distinct features at $\sim 1.1$, $1.4$ and $\sim 1.75~\mu m$ (indicated by boxes in the $1~\mathrm{Gyr}$ SED), consequence of the presence of CN and C2 (for details on these features, see \citet{aringer_et_al_2009} and \citet{loidl_et_al_2001}). This kind of stars is no longer present at ages older than $1.6~\mathrm{Gyr}$. For the MarS models, at older ages (middle panel), the absorption features, especially CO, become weaker, due to the diminishing presence of TP-AGB stars at older ages. We also see that the BaSS model at $1~\mathrm{Gyr}$ contains cooler TP-AGB stars than both GirS and MarS. However the extension to more evolved phases is present at all ages for the MarS models. This later phase is not seen in the isochrones for the GirS and BaSS models and therefore these models present a difference in spectral slope from MarS. In the lower panel, we see a similar trend for the MarS model SEDs, with the CO absorption features in the $K$ band becoming even weaker at $14~\mathrm{Gyr}$ since the stars at the TP-AGB phase are slightly warmer than at $7~\mathrm{Gyr}$. For the GirS and BaSS models, there is a more regular slope since the features become stronger towards the red. This is because at these wavelengths the TP-AGBs are slighter warmer at older ages and the main difference is the cooler temperatures of the MSTO and SGB. In the middle panel of this figure, we also observe the small differences due to the usage of different isochrones for this old populations.

Figure\,\ref{seds_age_metals} is similar to Figure\,\ref{seds_metal_ages}, but here we show the differences between the models at different metallicities. Telluric absorption features are present in all models at $\sim 1.40~\mu m$ and between the $H$ and $K$ bands. The main characteristic that can be seen from both figures is that the SEDs become redder as a function of metallicity (middle panel) since at higher metallicities, the opacities of the stars increase, and the temperatures are cooler. However, when comparing SEDs at solar and $0.2~\mathrm{dex}$ (lower panel), the differences diminish quite strongly showing that the models evolve very little after $\sim 5~\mathrm{Gyr}$. It is important to bear in mind that for solar and super solar models, the {\it IRTF spectral library} presents some limitations regarding the presence of cool bright giants.

In general, the differences seen in the SEDs at different ages and metallicities are expected to be a consequence of the different contributions of the RGB and TP-AGB stars to the spectra. RGB stars are old stars which have a stronger contribution at older ages and also as a function of redder wavelengths. Figures \,\ref{seds_metal_ages_J} to \,\ref{seds_age_metals_K} in the Appendix \,\ref{app01} are the same as Figures \,\ref{seds_metal_ages} and \,\ref{seds_age_metals} except zoomed into the wavelength ranges corresponding to $J$ ($1.04 - 1.44~\mu m$), $H$ ($1.46 - 1.84~\mu m$) and $K$ ($1.90 - 2.48~\mu m$) bands respectively. The spectra in these plots are given a constant offset in order to facilitate study. Thanks to the detailed analysis of atomic lines and molecular bands by \citet{rayner_et_al_2009} for the {\it IRTF spectral library} stars, and the compilation of \citet{ivanov_et_al_2004} in the $J$, $H$ and $K$ bands, we are able to easily identify several absorption features in our model SEDs using Figures\,\ref{seds_metal_ages_J} to \,\ref{seds_age_metals_K} where we observe the presence of features in the spectrum, i.e. the absorption line-strengths. We see that there are different trends for these line-strengths (see Figure\,\ref{full_seds}) as a function of age and metallicity. However, these trends can only be investigated in detail when the line-strength indices are calculated. In Section \,\ref{results_indices},  we present the trends of some line-strength indices in the $K$ band as a function of age.

\begin{figure*}
	\centering
	\includegraphics[width=\textwidth]{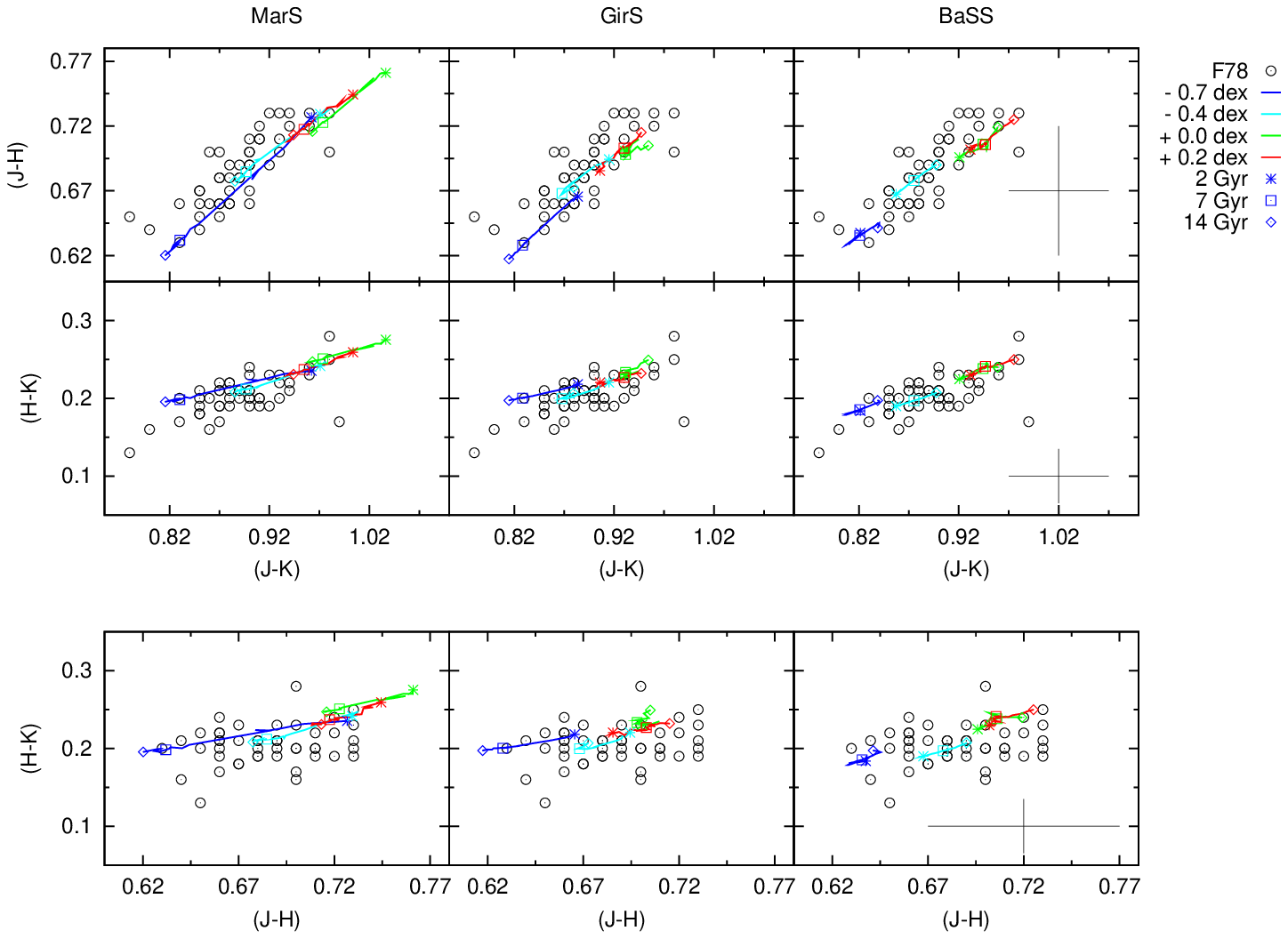}
	\caption{Colour-colour diagrams of our models at different metallicities and ages older than $2~\mathrm{Gyr}$, compared with observations of elliptical galaxies from \citet[][F78]{frogel_et_al_1978}. The error bars represent the typical uncertainties of the elliptical galaxy colours.}
	\label{ccd_obs}
\end{figure*}

      
\subsection{Integrated colours}
\label{results_colors}
\hspace{0.45cm}      
After obtaining the SEDs, we calculate the integrated colours of each model spectrum by integrating the spectral flux in the NIR colour bands using using the Vega spectrum from \citet{colina_et_al_1996} as a zero-point. We used the response curves of the $J$, $H$ and $K$ filters of the Johnson-Cousins-Glass photometric system given by \citet{bessel_et_al_1998}.

As explained in Section \,\ref{sec_iso}, we find a flux loss in the $H$ band which is taken into account also for the calculated magnitudes, increasing them by $0.0002\,\mathrm{mag}$. Additionally, our wavelength range does not completely cover the filter response curve for the $K$ band, which extends to $2.48~\mu m$. Following a procedure similar to that for the $H$ band flux loss, we calculate the magnitude in this band in the \citet{pickles_1998} stellar spectral flux library, with and without a complete filter response curve. With that, we obtain an average necessary gain of $\sim 0.07 \%$ or $0.0007\,\mathrm{mag}$. However, both factors are negligible making the integrated colours of our models directly comparable with observations and other authors' models.   
  
\begin{figure*}
	\centering
	\includegraphics[width=\textwidth]{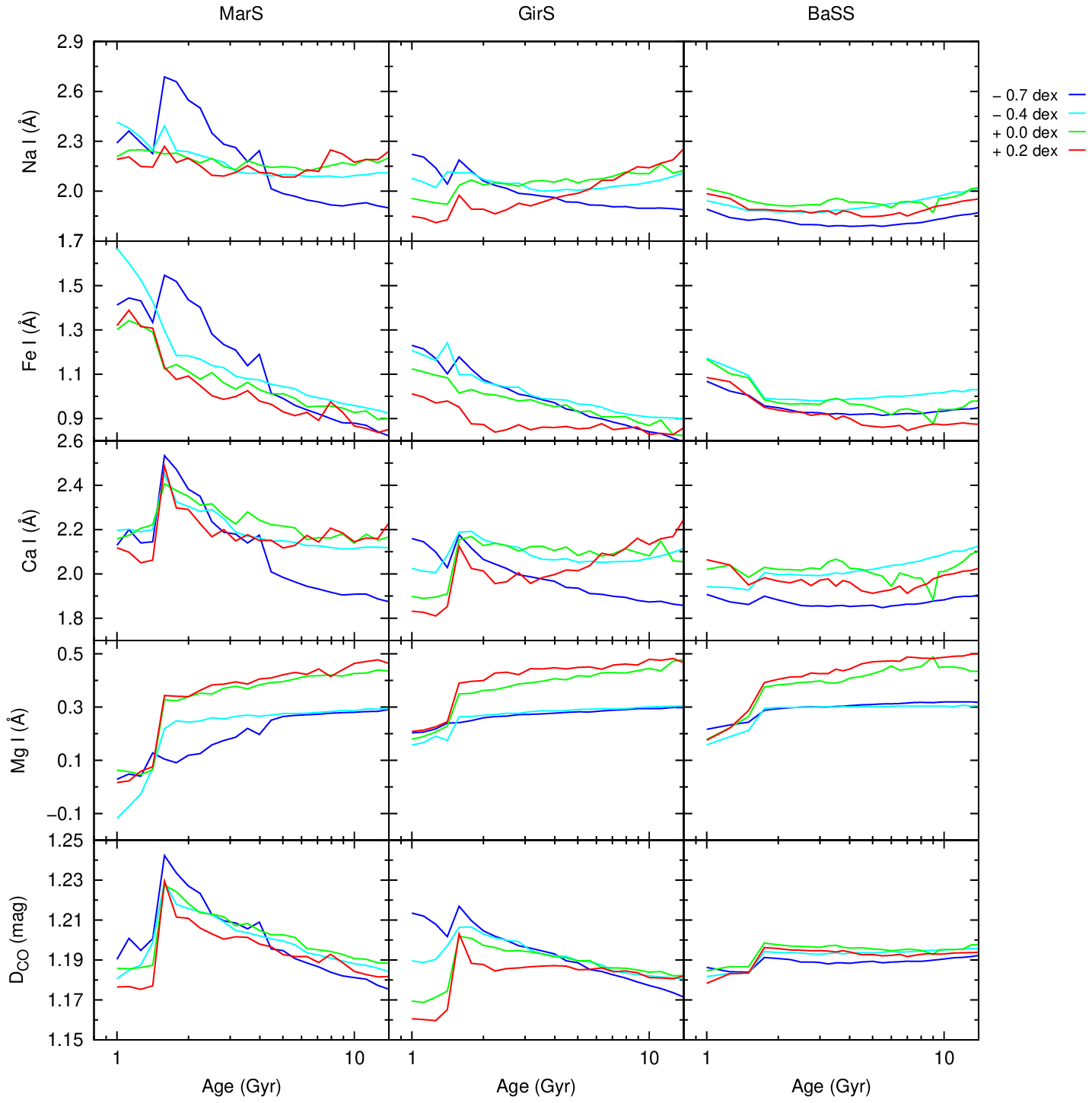}
	\caption{Comparison of different indices in the $K$ band as a function of age and metallicity, from models using different sets of isochrones. The models were convolved to a velocity dispersion of 350 $\mathrm{km\,s^{-1}}$ before measuring indices.}
	\label{age_indices}
\end{figure*}

Figure\,\ref{color_age} shows the behaviour of the integrated colours of the models as a function of age and compares them with observations of clusters in the Magellanic Clouds (MC) from \citet[][G04]{gonzalez-lopezlira_et_al_2004} and \citet[][P08]{pessev_et_al_2008}; and in the Milky Way from \citet[][C07]{cohen_et_al_2007}. For these three samples we chose those clusters with metallicities of $\sim-0.9~\mathrm{dex}$ or higher. It is worth mentioning that the ages and metallicities of the G08 sample were determined by creating "superclusters" by combining the photometry of the individual stars, comparing these with SSP models, and thereby obtaining average parameters, whereas the ages and metallicities of the P08 MC clusters were calculated by comparing individual clusters to SSP models. This approach of parameter determination presents a circular method when using this kind of data to determine the accuracy of other SSP models. For the C07 sample, we used ages determined by \citet{marin-franch_et_al_2009} when available and their average age of $13~\mathrm{Gyr}$ otherwise. All of our models cover the colour-colour range in $(J-K)$ and $(J-H)$ for the clusters within 2 sigma. It seems, however, that the MarS models are better able to reproduce the range in $(J-H)$ and $(H-K)$ colours of the individual clusters at younger ages. This is most likely due to the longer TP-AGB lifetimes in the M08 models. The effect of the AGB stars is even more evident at very low metallicities where the lifetimes of the AGB stars are considerably longer than for more metal-rich populations. Compared to the MarS models, the GirS and BaSS models do not include a detailed prescription of the AGB phase. In this figure we also notice the presence of the AGB stars at solar and higher metallicities because, in these models, due to their respective isochrones, the AGBs exist in these populations. For the BaSS models, the TP-AGBs phase seems to start at ages younger than $1~\mathrm{Gyr}$, which we currently cannot model given the limitations in parameter coverage of the {\it IRTF spectral library}. For older ages, the RGB stars are the main component of the light of these populations. The colours of these old populations show a linear behaviour (given the small range in colours) at constant metallicity. However, a point of interest is that for old populations, the RGB phase and the differences in temperature of the each model isochrones contribute to trend of these SSPs.

In Figure\,\ref{ccd_obs} we present $(J-K)$, $(J-H)$ and $(H-K)$ colour-colour diagrams of our models for different metallicities and ages older or equal than $2~\mathrm{Gyr}$, and compare them to observations of bright early-type galaxies by \citet[][F78]{frogel_et_al_1978}. Our three models are able to cover most of the colour range of this type of galaxy, with the MarS models matching the range in galaxy colours more closely. The presence of TP-AGBs and their extension to cooler temperatures at younger ages in the MarS models allows the colours to extend to the redder end of these galaxies. The age-metallicity degeneracy is present in the three sets of models 

In Tables\,\ref{MARSparameters} to\,\ref{BASSparameters} in the Appendix\,\ref{app02}, we present the integrated colours of each model set in the Johnson-Cousins-Glass photometric system given by \citet{bessel_et_al_1998}.

\begin{figure*}
	\centering
   	\includegraphics[width=\textwidth]{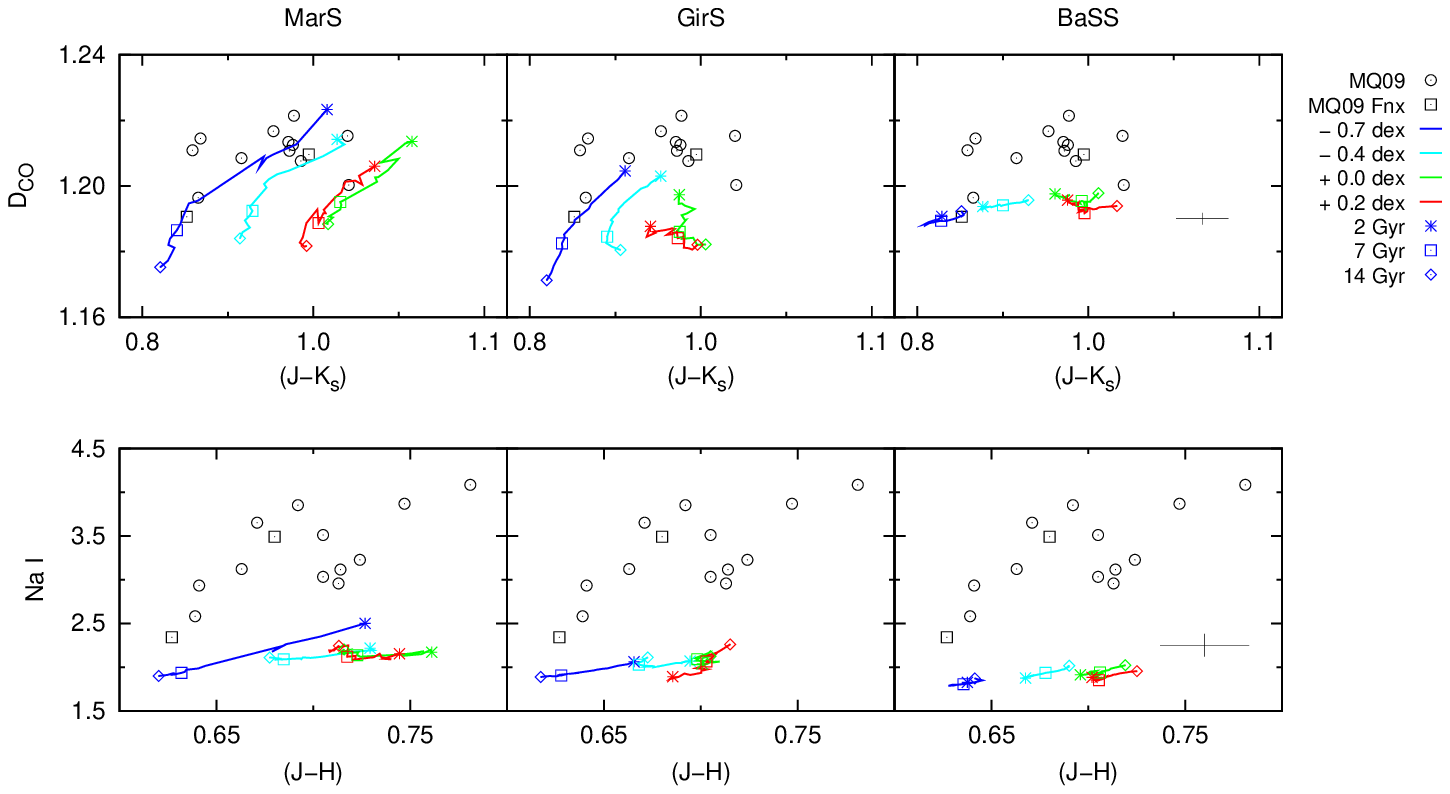}
	\caption{Behaviour of the molecular index D$_{CO}$ as a function of $(J-K_{s})$ (upper panels) and the atomic index Na I as a function of $(J-H)$ (lower panels) at different metallicities and ages older than $2~\mathrm{Gyr}$, compared with galaxies from \citet[][MQ09]{marmol-queralto_et_al_2009}. The models and the observations were convolved to a velocity dispersion of $350~\mathrm{km\,s^{-1}}$ before measuring indices. The error bars represent the typical uncertainties of the elliptical galaxy colours and indices}
	\label{color_indices}
\end{figure*}  

      
\subsection{Line-strength indices}
\label{results_indices}
\hspace{0.45cm}

Spectral features are another source of information that can be obtained from a model SED and compared with observations to determine stellar population properties. In this work, we focus on absorption line-strength indices in a region of the $K$ band. These indices are the atomic Na I, Ca I, Fe I and Mg I features, as well as the molecular CO feature. We apply the definitions of \citet{frogel_et_al_2001} for Na I and Ca I, \citet{silva_et_al_2008} for the Fe I doublet, $\,\mathrm{Fe~I = ( Fe~Ia + Fe~Ib ) / 2}$, and Mg I, and \citet{marmol-queralto_et_al_2008} for $D_{CO}$ (see Table\,\ref{tab_indices}).

Figure\,\ref{age_indices} presents the line-strength indices as a function of age of the models. As we can see, the indices follow a similar trend as the integrated colours, displaying the distinctive peak of the AGB phase in the MarS and GirS models. In the MarS models at older ages, the indices show an opposite behaviour as a function of age to the GirS and BaSS models. This is due to the AGB lifetimes, which cause these cool stars to dominate, hence the indices are higher than in the absence of these stars. All the indices are proportional to the mean effective temperature of the stars that compose the populations, except for Mg I since this feature is strongly driven by the stellar surface gravity \citep{viti_and_jones_1999} These trends are presented in detail for the {\it IRTF spectral library} in \citet{cesetti_et_al_2013}. Therefore, these indices would allow predictions of the stellar content in galaxies, especially to determine the presence of TP-AGBs. This can be seen by the characteristic peak of these stars in the indices at younger ages.      

We also present the indices $D_{CO}$ and Na I as a function of $(J-K_{s})$ and $(J-H)$ respectively for the three models in Figure\,\ref{color_indices}. The models are compared to observed elliptical galaxies from \citet[][hereafter MQ09]{marmol-queralto_et_al_2009}, of which 12 are field galaxies (circles) and two more belong to the Fornax cluster (squares). We obtained the indices from the spectra of these galaxies. The colours of models and the galaxies are compared in the 2MASS photometric system. To make a proper comparison, the model and the galaxy spectra were convolved to a uniform velocity dispersion of $350~\mathrm{km\,s^{-1}}$. For the galaxies, we took into account their instrumental resolution (7.2~\AA) and their intrinsic individual velocity dispersion, and for our models, the resolution in the $K$ band at $2.27~\mu m$ of 9.7~\AA.

Figure\,\ref{color_indices} shows that the MarS models best reproduce the range of $D_{CO}$ and colour of the observed galaxies, while GirS and BaSS are not able to cover the high $D_{CO}$ indices of part of the sample. This shows that the AGB population seems to be needed to reach the high observed $D_{CO}$ values. MQ09 have tested this by investigating indices in both the field and in the Fornax cluster. Fornax galaxies should have a smaller AGB fraction. This interpretation would agree with our models here. However, it is in principle also possible that the [C/Fe] or [O/Fe] abundance ratio is higher than solar, so that the observed $D_{CO}$ in all galaxies is higher than in the models. This is unlikely, however, for all galaxies since it would also be the case for the smallest galaxies of the sample, for which C/Fe from optical indices is not overabundant \citep{sanchez-blazquez_et_al_2006b}. Rather, it is likely that both [C/Fe] increases and the presence of AGB stars decreases with velocity dispersion of galaxies. Therefore, the high values and flat trend of $D_{CO}$ could be a combined effect of the increase oin[C/Fe] with increasing velocity dispersion mentioned above and the decreasing importance of the AGB phase with increasing metallicity seen in our models, coupled with the increasing metallicity of early-type galaxies with increasing velocity dispersion at fixed age \citep[e.g.][]{trager_et_al_2000a,sanchez-blazquez_et_al_2006b,trager_et_al_2008}. It will be interesting to test this by measuring the line-strength features of atomic C found in the $J$ and $H$ bands and their behaviour as a function of metallicity, and compare this with $D_{CO}$. {\bf Such comparisons would allow one to distinguish different episodes of star formation in a galaxy and even shed some light onto its chemical evolution.}

This figure additionally shows that none of the three models are able to reproduce Na I as a function of $(J-H)$. This could be due to the known [Na/Fe] overabundance seen in galaxies when Na indices are measured in the optical and NIR \citep[e.g.][]{jeong_et_al_2013}. In addition, the under-prediction of Na I by the models \citep{conroy_and_van_dokkum_2012} could be due to an IMF effect, specifically the presence of more dwarfs in the population. It is noteworthy that these galaxies span a large range in velocity dispersion between $\sim90$ up to $\sim310~\mathrm{km\,s^{-1}}$ \citep{marmol-queralto_et_al_2009}. Given recent work that the IMF is closely linked to their velocity dispersions (mass) \citep{capellari_et_al_2012}, a comparison with models with different IMFs and $\alpha$-enhancement conditions would be appropriate. In our following paper (Paper III), we will address this issue in more detail where we will also present the behaviour of other indices in the $K$ band for these galaxies.

\begin{table}
	\begin{center}
	\caption[]{Index definitions used in this work.}
         \label{tab_indices}
         	\resizebox{0.5\textwidth}{!}{
         \footnotesize
          \begin{tabular}{lcccc}\hline
          	 			& 		  & {\bf Bandpass (\AA)} &		          \\
		{\bf Index}		& {\bf Blue} & {\bf Central} 		 & {\bf Red}         \\ \hline
	 	\hline
        		Na I	$^{a}$	& $21910 - 21966$ & $22040 - 22107$ & $22125 - 22170$ \\
		\hline
        		Ca I	$^{a}$	& $22450 - 22560$ & $22577 - 22692$ & $22700 - 22720$ \\
		\hline
        		Fe Ia	 $^{b}$	& $22133 - 22176$ & $22251 - 22332$ & $22465 - 22560$ \\
		\hline
        		Fe Ib	 $^{b}$	& $22133 - 22176$ & $22369 - 22435$ & $22465 - 22560$ \\
		\hline
        		Mg I	$^{b}$	& $22715 - 22755$ & $22790 - 22850$ & $22850 - 22874$ \\
		\hline
		D$_{CO}$	 $^{c}$& $22872 - 22925$ & $22930 - 22983$ &				 \\
				& $22710 - 22770$ &                                 & \\
		\hline
         \end{tabular}
         }
         \end{center}
         \medskip
         
         Definitions by ($^{a}$) \citet{frogel_et_al_2001}, ($^{b}$) \citet{silva_et_al_2008} and ($^{c}$) \citet{marmol-queralto_et_al_2008}.
\end{table}

In Tables\,\ref{MARSparameters} to\,\ref{BASSparameters} in the Appendix\,\ref{app02}, we present the line-strength indices of each model set, at velocity dispersion of $350~\mathrm{km\,s^{-1}}$.

\begin{figure*}
	\centering
	\includegraphics[width=\textwidth]{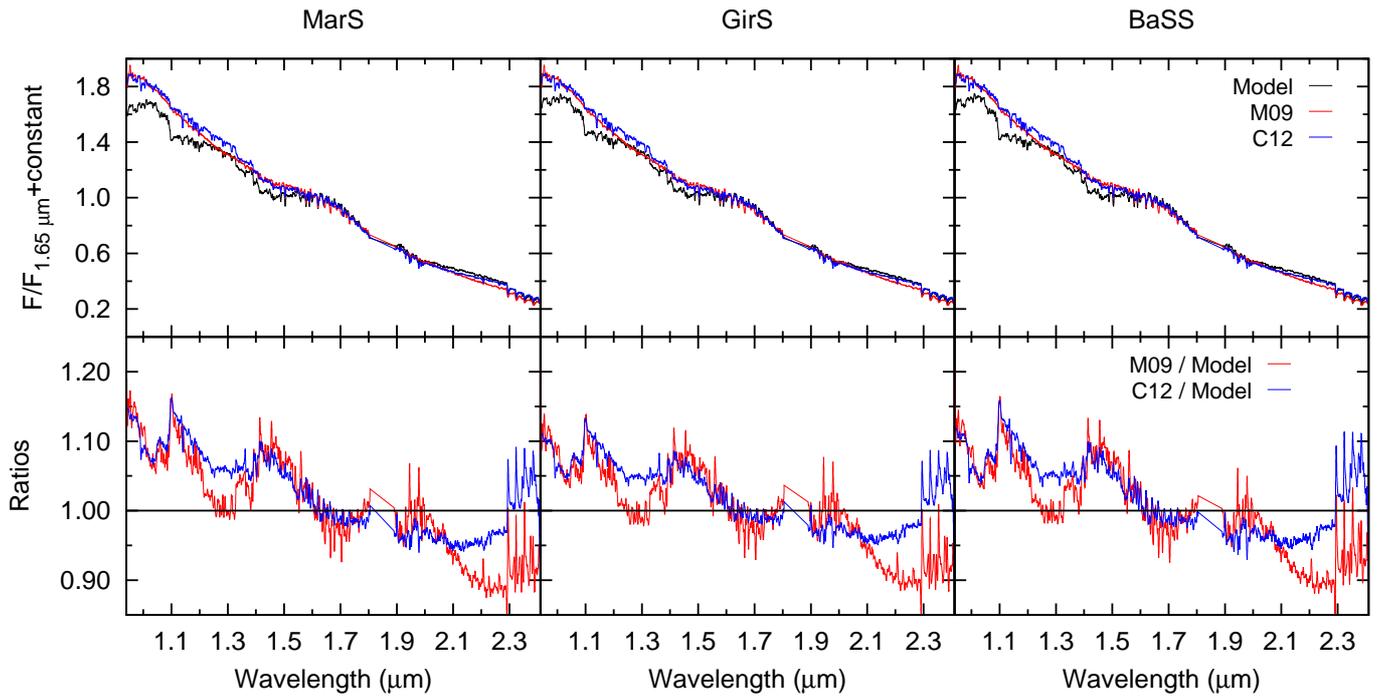}
	\caption{Comparison of our three models with those of \citet[][C12]{conroy_and_van_dokkum_2012} and \citet[][M09]{maraston_et_al_2009b}, at solar metallicity and $11~\mathrm{Gyr}$. The spectra are normalised to unity at $1.65~\mu m$ and convolved to 20~\AA.}
	\label{seds_others}
\end{figure*}

\begin{figure*}
	\centering
	\includegraphics[width=\textwidth]{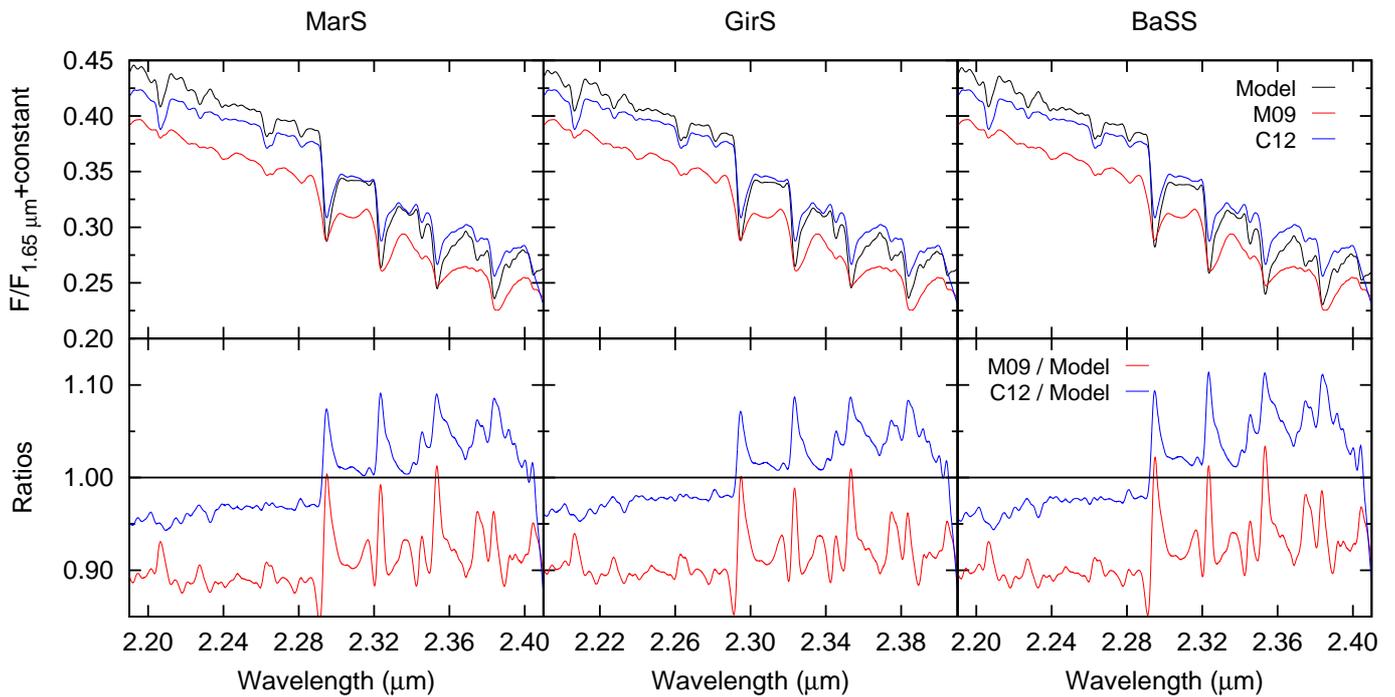}
	\caption{Zoom in of a section of the $K$ band ($2.19 - 2.42~\mu m$) for the comparison of our three models with those of \citet[][C12]{conroy_and_van_dokkum_2012} and \citet[][M09]{maraston_et_al_2009b}, at solar metallicity and $11~\mathrm{Gyr}$. The spectra are normalised to unity at $1.65~\mu m$ and convolved to 20~\AA.}
	\label{seds_others_Kband}
\end{figure*}

\begin{figure*}
	\centering
	\includegraphics[width=\textwidth]{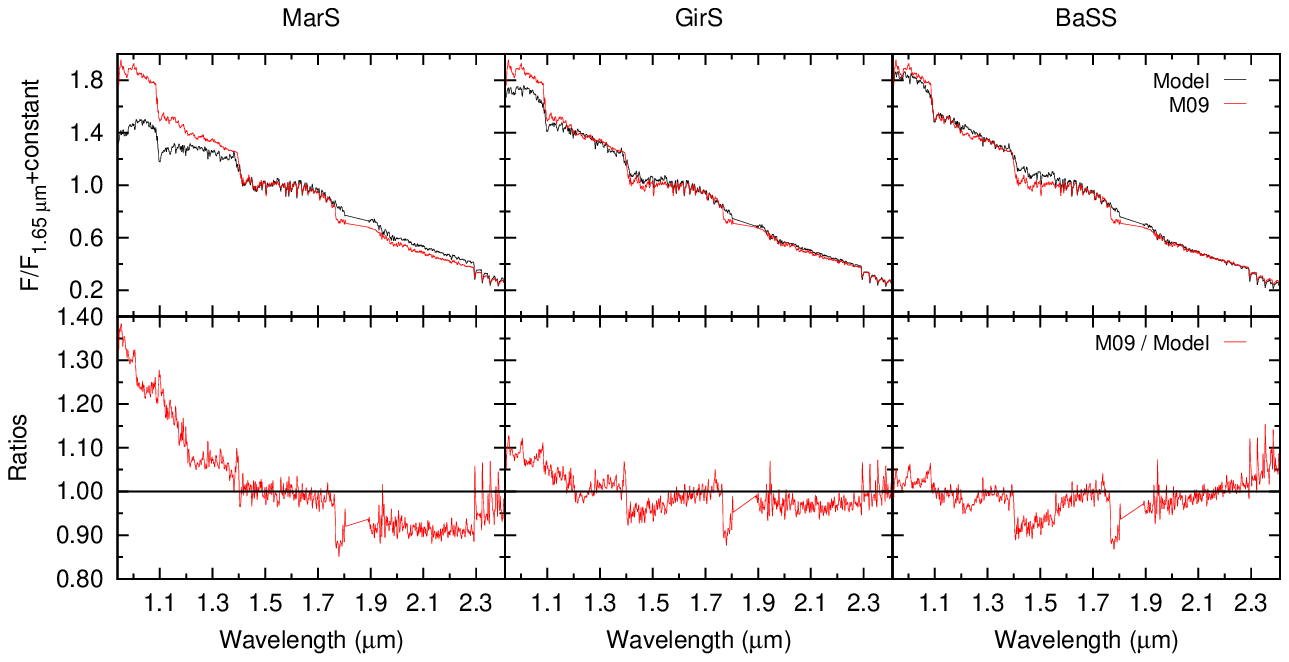}
	\caption{Comparison of our three models with the models of \citet[][M09]{maraston_et_al_2009b}, at solar metallicity and $1.5~\mathrm{Gyr}$. The spectra are normalised to unity at $1.65~\mu m$ and convolved to 20~\AA.}
	\label{seds_maraston}
\end{figure*}


\section{Comparisons with other authors}
\label{results_authors}
\hspace{0.45cm}      

Other authors have also made SSP models in the NIR range \citep[e.g.][]{conroy_and_van_dokkum_2012,maraston_et_al_2009b} using combinations of different empirical and theoretical stellar libraries. We compare our models with the models of \citet[][hereafter C12]{conroy_and_van_dokkum_2012} and \citet[][hereafter M09]{maraston_et_al_2009b}. The C12 models are partially based on the {\it IRTF spectral library} but they only take into account the stars that are also found in the MILES library in order to complement the spectra from the optical to the NIR range. For this sample they assume solar metallicity which in principle is valid. However, in our models we determined the stellar parameters of this library (Paper I) allowing us to make use of all the IRTF stars. The C12 models use two sets of isochrones depending on the stellar evolution stage: for the AGBs they use the M08 isochrones and for the RGBs the Dartmouth isochrones \citep{dotter_et_al_2008}. On the other hand, the M09 models use a theoretical library \citep[BaSeL models from][]{lejeune_et_al_1997,lejeune_et_al_1998,westera_et_al_2002} with a resolution of 20~\AA, complemented with the \citet{lancon_and_wood_2000} library of cool stars at $R \sim 1000$. For the isochrones, M09 uses a different treatment, especially for the AGB phase, where the fuel consumption theorem is used \citep{renzini_1981}. As we previously mentioned, theoretical libraries have limitations when reproducing molecular features \citep{martins_and_coelho_2007}. 

In Figure\,\ref{seds_others} we compare the SEDs at solar metallicity and $11~\mathrm{Gyr}$ of our models, C12 and M09. Since our models and C12 share, for the same wavelength range, the spectra of the {\it IRTF spectral library}, we took into account the resolution of 9.7~\AA in the $K$ band which we measured (Paper I), and in order to have a better comparison, we convolved all the model spectra to the lowest resolution of the M09 models. To improve the comparison, we also normalised the spectra to unity at $1.65~\mu m$. Overall, our models show the same continuum slope (within $10\%$) as the models of the other authors. When comparing in detail with C12 we see a change in slope at $\sim1.4~\mu m$ and $\sim2.2~\mu m$ across the wavelength range and comparing with M09, our models present an overall flatter slope. 

We analyse in detail the absorption spectral features of Na I, Fe I, Ca I, Ma I and CO lines found in a section of the $K$ band, between $2.19$ and $2.42~\mu m$ (see Table\,\ref{tab_indices}) in Figure\,\ref{seds_others_Kband}. The CO features in the M09 and C12 models are shallower than in our models, which reflects a smaller contribution of AGB/RGB stars when compared with our models. When doing this analysis, we noticed a shift in the lines of the M09 models. This could be due to either the combination of different resolutions or of air-vacuum wavelengths of the two stellar libraries.

In Figure\,\ref{seds_maraston} we compared the models by M09 at young age ($1.5~\mathrm{Gyr}$) with our three models. The main difference between our models and M09 is the different prescriptions used for creating a population. The fuel-consumption theorem and the isochrone treatment in the M09 models produce a higher number of C-stars in young populations which can be seen from the depressions given by these stars at around $1.15~\mu m$, $1.45~\mu m$ and $1.75~\mu m$. 

\begin{figure}
	\centering
   	\includegraphics[width=\columnwidth]{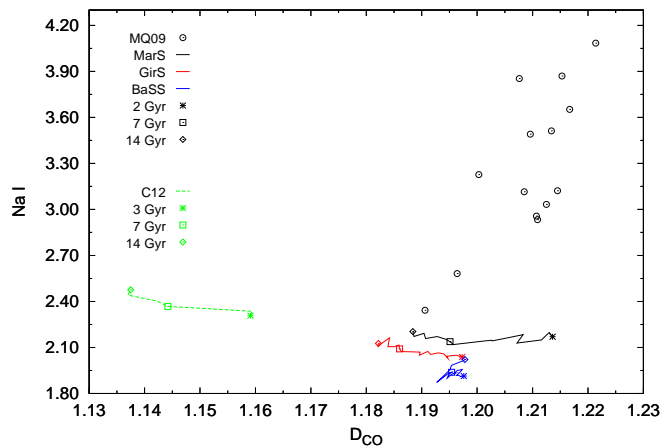}
	\caption{Comparison of line-strength indices Na I and $D_{CO}$ for our three models (MarS, GirS and BaSS) at solar metallicity with \citet[][C12]{conroy_and_van_dokkum_2012}. All models were convolved to a velocity dispersion of $350~\mathrm{km\,s^{-1}}$ before measuring indices.}
	\label{index_index}
\end{figure}

We also compared the line-strength indices in the $K$ band for Na I and $D_{CO}$ of our three models with C12, at solar metallicity. This index-index comparison is presented in Figure\,\ref{index_index}, which shows that our models have stronger $D_{CO}$ than the C12 models. This was already seen in the SED comparison, where the CO features were shallower in C12 than in our models. CO is strongly related to the presence of cooler stars in a population. Since the SSP models are related to how the isochrones populate different stellar phases, our models using the three sets of isochrones have a strong contribution of cool and AGB stars that allows us to reproduce the CO line strengths of local elliptical galaxies, as seen in Figure\,\ref{color_indices}. In contrast, the C12 models display a smaller contribution of AGB/RGB stars than observed in our models.

\begin{figure}
	\centering
   	\includegraphics[width=\columnwidth]{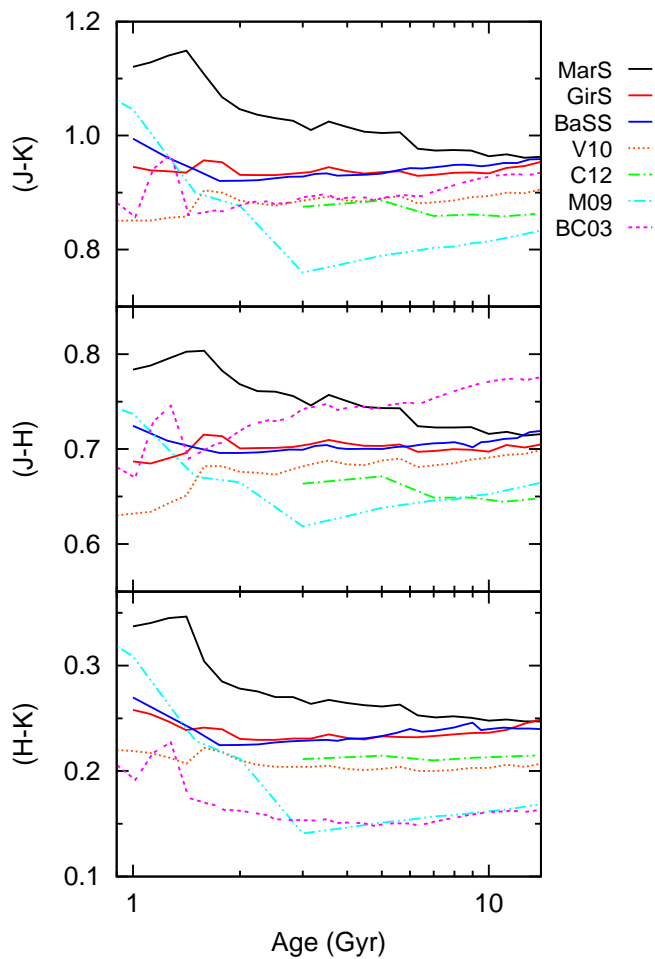}
	\caption{Comparison of integrated colours for our three models (MarS, GirS and BaSS) at solar metallicity with literature values: \citet[][B03]{bruzual_and_charlot_2003}, \citet[][V10]{vazdekis_et_al_2010}, \citet[][C12]{conroy_and_van_dokkum_2012} and \citet[][M09]{maraston_et_al_2009b}.}
	\label{authors}
\end{figure}

We made a comparison of the integrated colours of our models at solar metallicity to the available literature values of other authors such as \citet[][BC03]{bruzual_and_charlot_2003} and \citet[][V10]{vazdekis_et_al_2010}, as well as C12 and M09. The BC03 models are based on the G00 isochrones and the BaSeL theoretical spectral library. The integrated colours of V10 come from photometric predictions based on the transformations of \citet{alonso_et_al_1996,alonso_et_al_1999} for the G00 isochrones. In Figure\,\ref{authors}, we present the colours of our three models MarS, GirS and BaSS, and compare them with BC03, C12, M09 and V10. When comparing with other authors, our models follow the general trend. However, there is a large scatter in the colours of published models; for example BC03 and M09 have the bluest $(H-K)$ colours and are not in good agreement with giant ellipticals. We notice that at younger ages ($1 - 2~\mathrm{Gyr}$), the integrated colours of our MarS models are redder than most models. This is due to the treatment of TP-AGB stars by the M08 isochrones, which peak around these ages. As expected given the similar ingredients, our GirS and the V10 models behave quite similarly. Our BaSS model for $(J-K)$ also behaves quite similarly to V10 and for intermediate ages, to BC03. 

\section{Summary and final remarks}
\hspace{0.45cm}

In this series of papers,  we aim to to provide an improved tool for stellar population studies in the NIR range, primarily for the $J$, $H$ and $K$ bands. This wavelength coverage is strongly influenced by cool late type stars (e.g. AGB and RGB stars) which are relevant for a diverse age-range of stellar populations, including early-type galaxies (Section \,\ref{models_synthesis}). We use a single empirical stellar library with a trustworthy flux calibration (Section \ref{sec_irtf}) with homogenous stellar parameters (Paper I) and empirical transformations from the theoretical to the observational plane (Section \ref{sec_iso}). Therefore, our models are the first at intermediate resolution purely based on empirical spectra in the NIR range. The comparisons presented here show the power of our models for the analysis of old stellar populations like early-type galaxies.

In this work we present Single Stellar Population models synthesised with the {\it IRTF spectral library}, for ages from $1$ to $14~\mathrm{Gyr}$ and $[Z/Z_{\odot}]$ from $-0.70$ to $0.20~\mathrm{dex}$, over a wavelength range from $0.94$ to $2.41~\mu m$. 

By using three different sets of isochrones, we can see the relevance of different prescriptions for stellar evolution and their influence on the SEDs, integrated colours and indices. We have shown that the choice of isochrones is very important in determining the output for young ages, where the AGB dominates. 

The colour-colour trends that our models show are a good match to the colours of elliptical galaxies. We also compare indices of observed galaxies by smoothing the SEDs of our models and the observations to the same velocity dispersion (Figure\,\ref{color_indices}). Our models reproduce the $D_{CO}$ index of elliptical galaxies, giving confidence to the predictive power of our models. Our model SEDs compare well with other models in the literature, taking into account that detailed predictions for line strengths in this wavelength region in the literature are very scarce. 

The models presented in this work use a Salpeter IMF (see Section \,\ref{sec_imf}). Nonetheless, we are aware that even though recent studies provide evidence that the IMF is largely invariant throughout the Local Group \citep[e.g.][and references therein]{kroupa_2012} this may not apply outside of it, especially for elliptical galaxies \citep[e.g.][]{capellari_et_al_2012}. We will make a deeper analysis of the impact of different types of IMFs on stellar population studies in a future publication. 

Our models can be used to study the SEDs of galaxies in a versatile way with full-spectrum fitting or focusing on selected features. Our models, based on a empirical stellar spectral library with moderate resolution, reproduce the NIR observations of clusters and galaxies, as desired. In Paper III, we will use both approaches to analyse the spectra of a sample of field and cluster galaxies, and derive their stellar population properties such as ages and metallicities. 

\section*{Acknowledgments}
\hspace{0.45cm}
The authors acknowledge the usage of the SIMBAD data base and VizieR catalogue access tool (both operated at CDS, Strasbourg, France). The authors would like to thank Charlie Conroy for providing his models and E. M{\'{a}}rmol-Queral{\'{o}} for the sample spectra. Additionally, we would like to thank J. Falc{\'{o}}n-Barroso and M. Koleva for their help with the characterisation of the stellar library and their useful discussions. SMG thanks Ariane Lan\c con for bringing to her attention to the role of C-stars in SSP modelling. SMG also thanks T. de Boer and the Institute of Astronomy of the University of Cambridge for support during her stay. AV acknowledges the support by the DAGAL collaboration and the Programa Nacional del Astronom{\'{i}}a y Astrof{\'{i}}sica of the Spanish Ministry of Science and Innovation under grant AYA2010$-$21322$-$C03$-$02. 


\bibliographystyle{aa}
\bibliography{references}

\onecolumn
\clearpage
\appendix

\section[]{Zoom in to individual bands}
\label{app01}
\begin{figure*}[!h]
	\centering
	\includegraphics[width=0.85\textwidth]{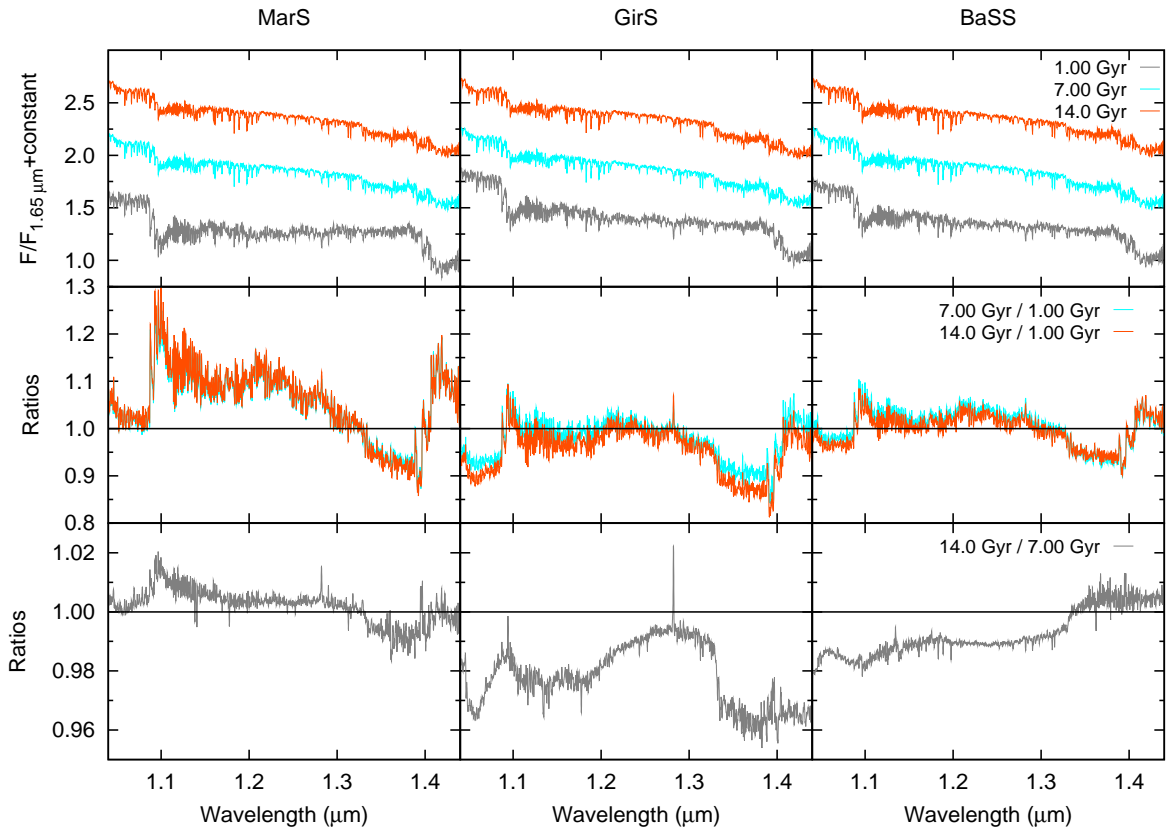}
	\caption{Same as Figure \,\ref{seds_metal_ages} except over the J band (1.04 $-$ 1.44 $\mu$m).}
	\label{seds_metal_ages_J}
\end{figure*}

\begin{figure*}[!htb]
	\centering
	\includegraphics[width=0.85\textwidth]{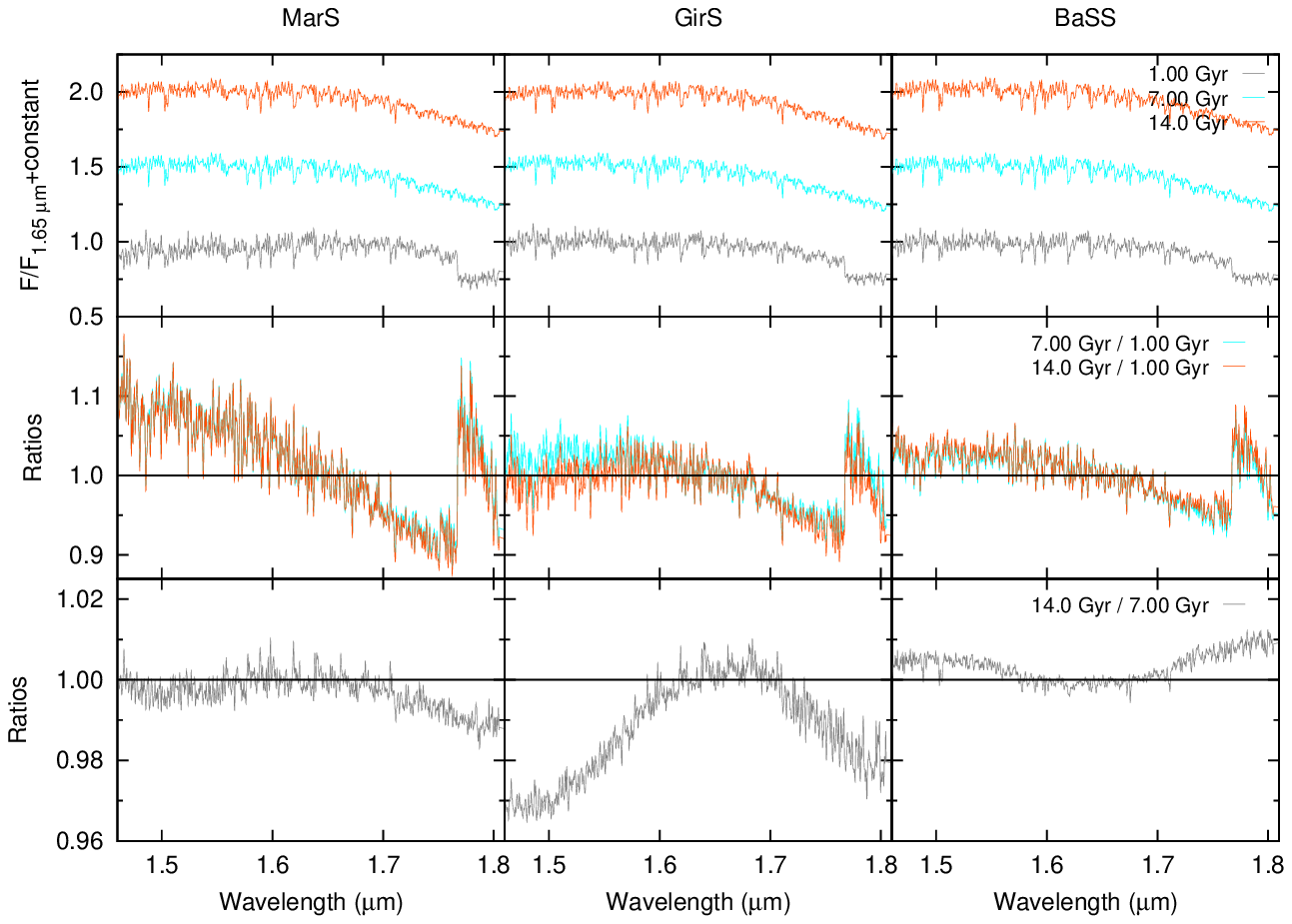}
	\caption{Same as Figure \,\ref{seds_metal_ages} expect over the H band (1.46 $-$ 1.84 $\mu$m).}
	\label{seds_metal_ages_H}
\end{figure*}

\begin{figure*}[!htb]
	\centering
	\includegraphics[width=0.85\textwidth]{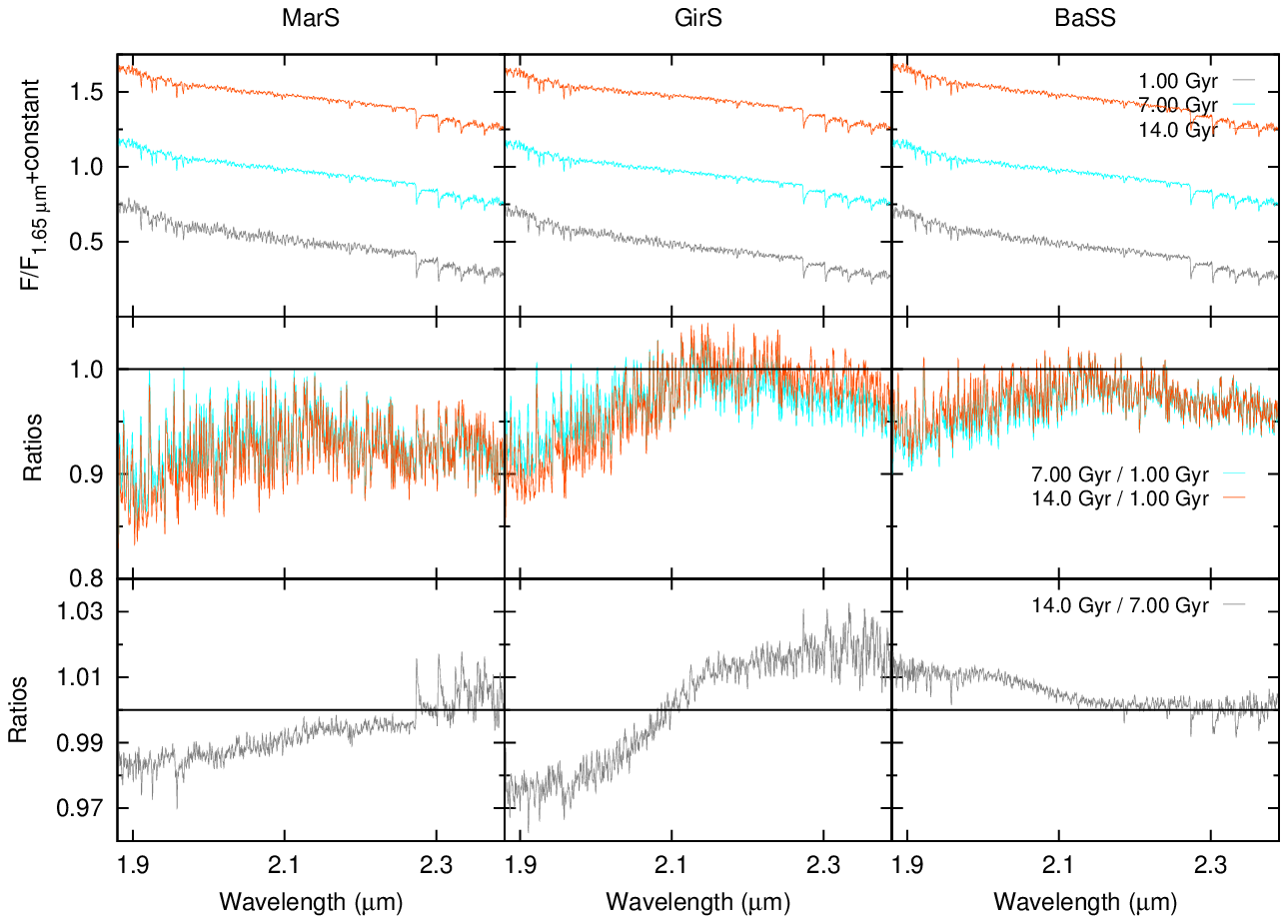}
	\caption{Same as Figure \,\ref{seds_metal_ages} expect over the K band (1.90 $-$ 2.48 $\mu$m).}
	\label{seds_metal_ages_K}
\end{figure*}

\begin{figure*}[!htb]
	\centering
	\includegraphics[width=0.85\textwidth]{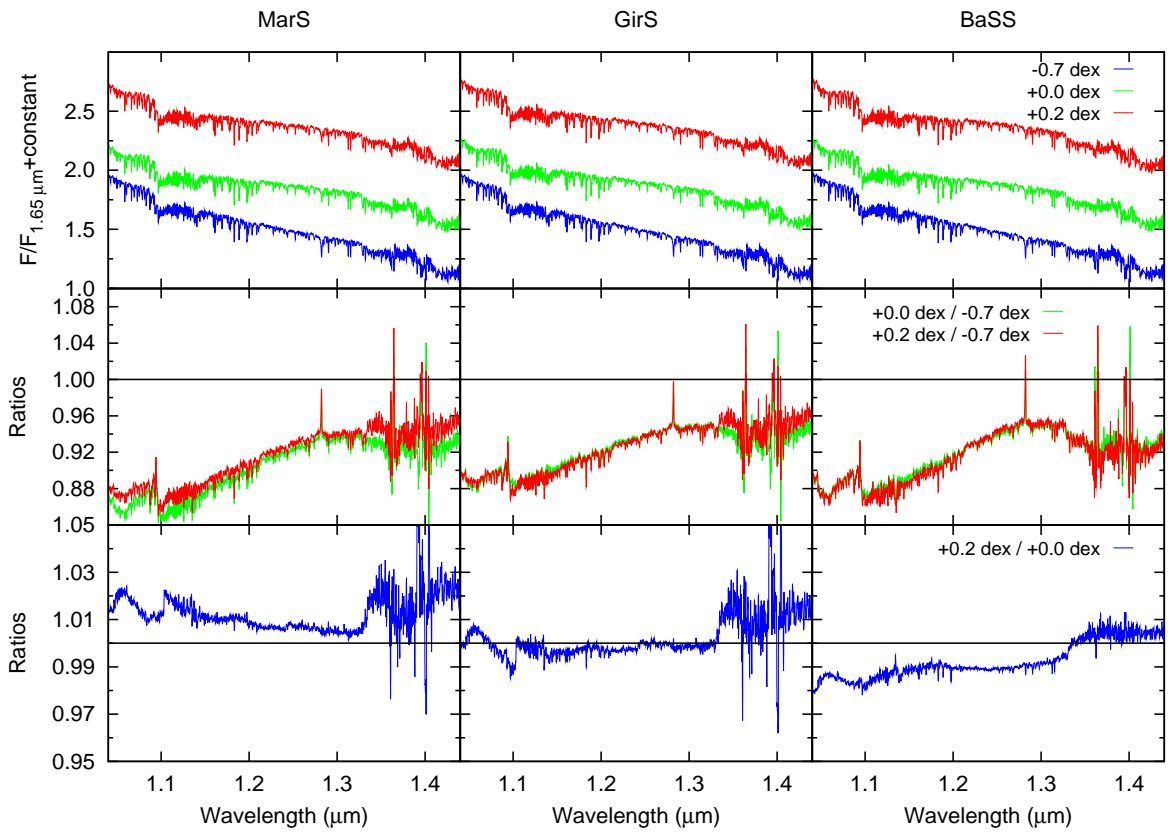}
	\caption{Same as Figure \,\ref{seds_age_metals} expect over the J band (1.04 $-$ 1.44 $\mu$m).}
	\label{seds_age_metals_J}
\end{figure*}

\begin{figure*}[!htb]
	\centering
	\includegraphics[width=0.85\textwidth]{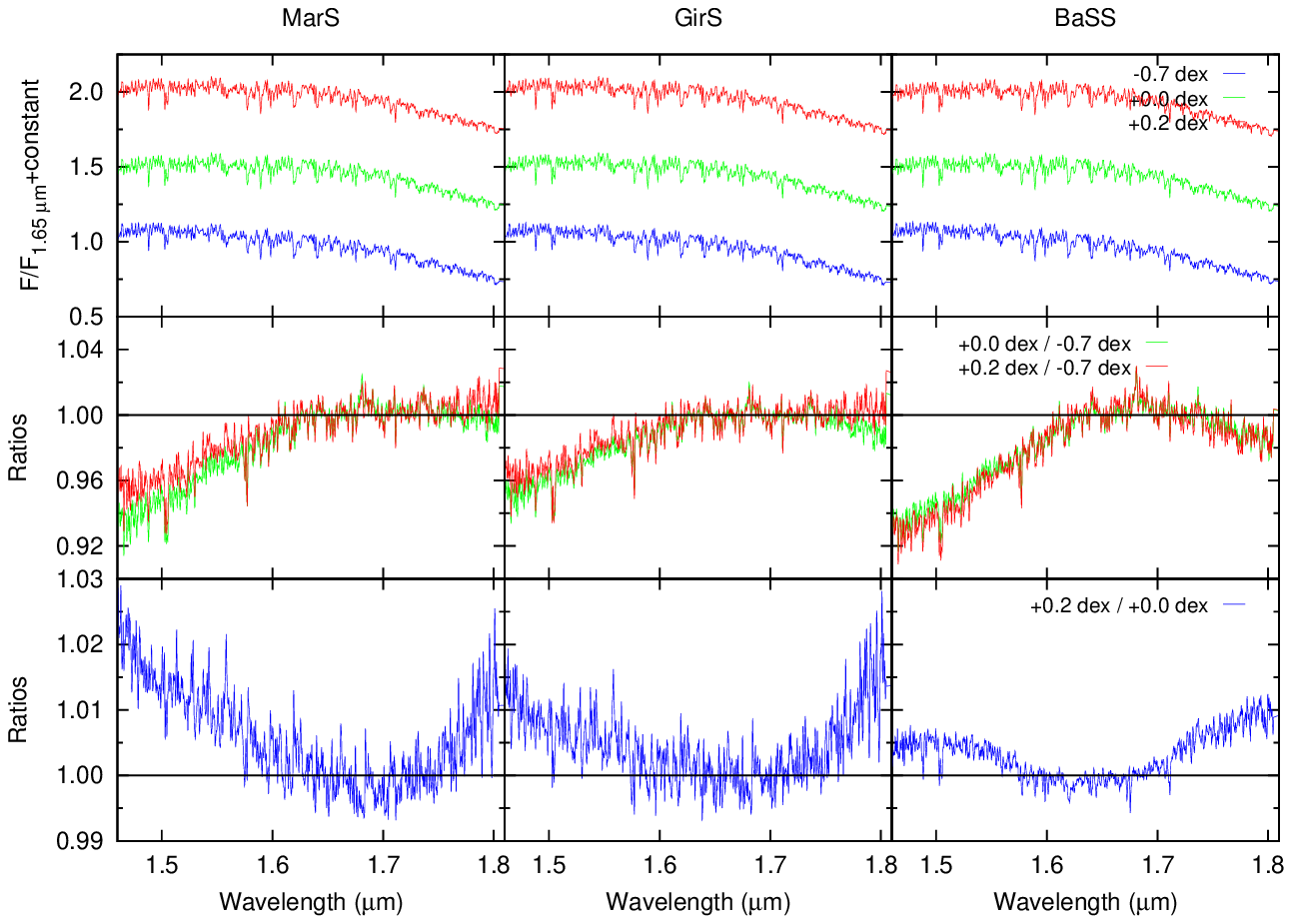}
	\caption{Same as Figure \,\ref{seds_age_metals} expect over the H band (1.46 $-$ 1.84 $\mu$m).}
	\label{seds_age_metals_H}
\end{figure*}

\begin{figure*}[!htb]
	\centering
	\includegraphics[width=0.85\textwidth]{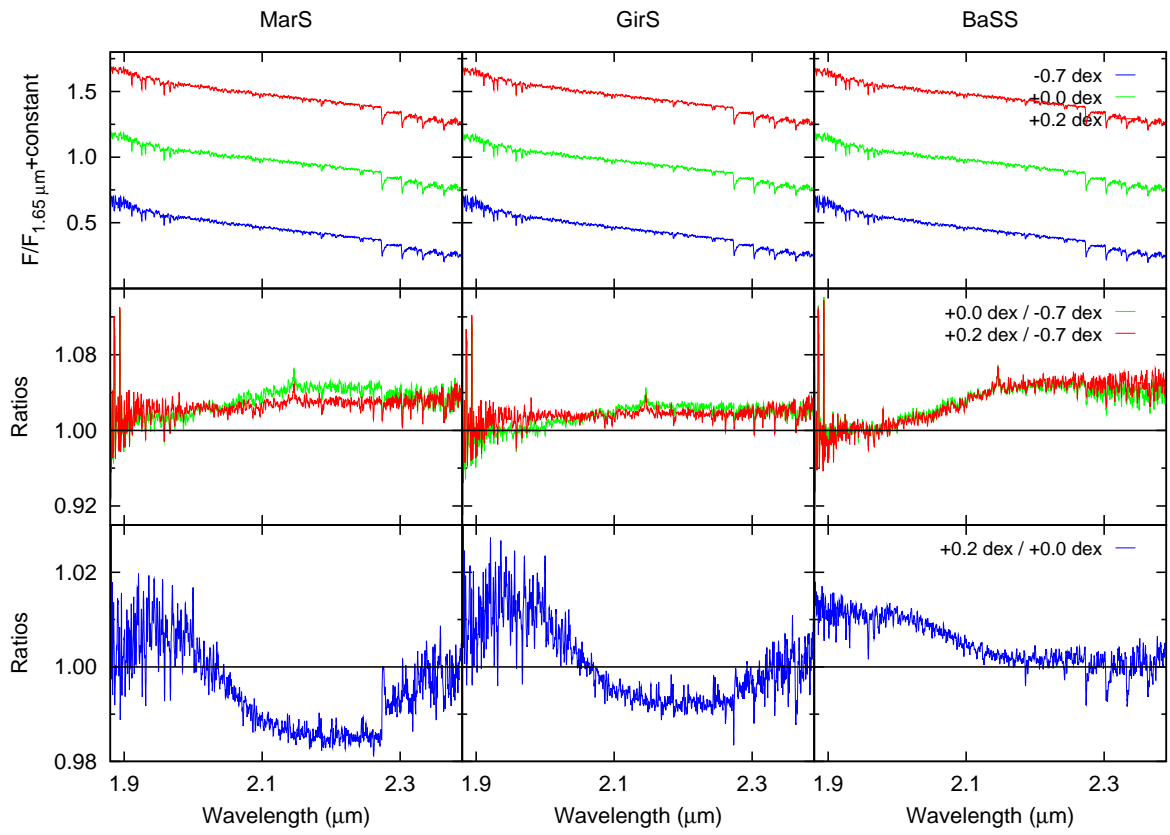}
	\caption{Same as Figure \,\ref{seds_age_metals} expect over the K band (1.90 $-$ 2.48 $\mu$m).}
	\label{seds_age_metals_K}
\end{figure*}

In this section, we present the Spectral Energy Distribution of our Single Stellar Population models over the $J$, $H$ and $K$ bands. Figure~\ref{seds_metal_ages_J} to~\ref{seds_metal_ages_K} present the SEDs at solar metallicity and different ages. Figure~\ref{seds_age_metals_J} to~\ref{seds_age_metals_K} present the SEDs at solar metallicity and different ages. \\

\clearpage

\section[]{Integrated colours and line-strength indices from our models}
\label{app02}
These are the integrated colours and the line strength indices from the Spectral Energy Distributions of the our Single Stellar Population. The SEDs were convolved to a velocity dispersion of $350~\mathrm{km\,s^{-1}}$ before calculating indices.

\begin{footnotesize}
\begin{longtable}{rrrrrrrrrr}
 \caption{Integrated colours and abundances for the MarS models as a function of age and metallicity}
 \label{MARSparameters}	\\
 \hline
 \hline
 {\bf Age ($\mathrm{Gyr}$)} & {\bf [Z/Z$_{\odot}$]} & {\bf (J-K)} & {\bf (J-H)} & {\bf (H-K)} & {\bf Na I (\AA)} & {\bf Fe I (\AA)} & {\bf Ca I (\AA)} & {\bf Mg I (\AA)} & {\bf D$_{\mathrm{CO}}$ (mag)} \\ 
 \hline
 \hline
 \endfirsthead
 \hline
 \hline
 \endlastfoot
 \multicolumn{10}{c}{\tablename\ \thetable\ - Continued from previous page} \\
 \multicolumn{10}{c}{ } \\
 \hline
 \hline
 {\bf Age ($\mathrm{Gyr}$)} & {\bf [Z/Z$_{\odot}$]} & {\bf (J-K)} & {\bf (J-H)} & {\bf (H-K)} & {\bf Na I (\AA)} & {\bf Fe I (\AA)} & {\bf Ca I (\AA)} & {\bf Mg I (\AA)} & {\bf D$_{\mathrm{CO}}$ (mag)} \\ 
 \hline
 \hline
 \endhead
 \hline
 \hline
 \endfoot
1.00  & $-$0.70 & 0.987 & 0.700 & 0.287 & 2.290 & 1.412 & 2.129 & 0.028 & 1.190 \\
1.12  & $-$0.70 & 1.004 & 0.716 & 0.288 & 2.361 & 1.443 & 2.200 & 0.048 & 1.200 \\
1.25  & $-$0.70 & 1.014 & 0.723 & 0.290 & 2.290 & 1.430 & 2.139 & 0.040 & 1.194 \\
1.41  & $-$0.70 & 0.996 & 0.732 & 0.264 & 2.225 & 1.333 & 2.145 & 0.127 & 1.200 \\
1.58  & $-$0.70 & 1.054 & 0.787 & 0.266 & 2.686 & 1.546 & 2.533 & 0.104 & 1.242 \\
1.77  & $-$0.70 & 1.012 & 0.755 & 0.256 & 2.657 & 1.518 & 2.471 & 0.090 & 1.233 \\
1.99  & $-$0.70 & 0.982 & 0.737 & 0.245 & 2.548 & 1.436 & 2.382 & 0.118 & 1.227 \\
2.23  & $-$0.70 & 0.962 & 0.726 & 0.235 & 2.500 & 1.401 & 2.349 & 0.125 & 1.223 \\
2.51  & $-$0.70 & 0.935 & 0.703 & 0.231 & 2.349 & 1.281 & 2.235 & 0.157 & 1.212 \\
2.81  & $-$0.70 & 0.914 & 0.690 & 0.224 & 2.282 & 1.233 & 2.188 & 0.174 & 1.209 \\
3.16  & $-$0.70 & 0.908 & 0.685 & 0.223 & 2.262 & 1.208 & 2.179 & 0.186 & 1.208 \\
3.54  & $-$0.70 & 0.902 & 0.678 & 0.223 & 2.176 & 1.138 & 2.139 & 0.220 & 1.205 \\
3.98  & $-$0.70 & 0.906 & 0.684 & 0.222 & 2.243 & 1.189 & 2.174 & 0.197 & 1.209 \\
4.46  & $-$0.70 & 0.848 & 0.644 & 0.204 & 2.014 & 1.014 & 2.009 & 0.250 & 1.195 \\
5.01  & $-$0.70 & 0.840 & 0.640 & 0.200 & 1.985 & 0.990 & 1.984 & 0.264 & 1.194 \\
5.62  & $-$0.70 & 0.835 & 0.634 & 0.201 & 1.969 & 0.959 & 1.963 & 0.268 & 1.190 \\
6.31  & $-$0.70 & 0.833 & 0.633 & 0.199 & 1.946 & 0.939 & 1.944 & 0.270 & 1.188 \\
7.08  & $-$0.70 & 0.830 & 0.631 & 0.198 & 1.935 & 0.921 & 1.931 & 0.273 & 1.186 \\
7.94  & $-$0.70 & 0.824 & 0.628 & 0.195 & 1.917 & 0.900 & 1.917 & 0.276 & 1.183 \\
8.91  & $-$0.70 & 0.822 & 0.626 & 0.196 & 1.911 & 0.880 & 1.904 & 0.278 & 1.182 \\
10.00 & $-$0.70 & 0.828 & 0.628 & 0.199 & 1.922 & 0.879 & 1.908 & 0.280 & 1.181 \\
11.22 & $-$0.70 & 0.827 & 0.627 & 0.199 & 1.930 & 0.869 & 1.909 & 0.282 & 1.180 \\
12.59 & $-$0.70 & 0.822 & 0.624 & 0.197 & 1.910 & 0.840 & 1.887 & 0.284 & 1.177 \\
14.13 & $-$0.70 & 0.815 & 0.620 & 0.195 & 1.898 & 0.820 & 1.873 & 0.291 & 1.175 \\
 & & & & & & & & & \\
1.00  & $-$0.40 & 1.170 & 0.804 & 0.365 & 2.414 & 1.668 & 2.196 & $-$0.117 & 1.180 \\
1.12  & $-$0.40 & 1.134 & 0.787 & 0.346 & 2.379 & 1.600 & 2.201 & $-$0.073 & 1.184 \\
1.25  & $-$0.40 & 1.110 & 0.781 & 0.328 & 2.323 & 1.526 & 2.189 &-0.026 & 1.187 \\
1.41  & $-$0.40 & 1.057 & 0.775 & 0.282 & 2.247 & 1.427 & 2.199 & 0.072 & 1.198 \\
1.58  & $-$0.40 & 1.044 & 0.773 & 0.270 & 2.391 & 1.297 & 2.450 & 0.219 & 1.229 \\
1.77  & $-$0.40 & 0.998 & 0.743 & 0.254 & 2.244 & 1.184 & 2.324 & 0.248 & 1.217 \\
1.99  & $-$0.40 & 0.986 & 0.736 & 0.249 & 2.237 & 1.183 & 2.303 & 0.243 & 1.215 \\
2.23  & $-$0.40 & 0.970 & 0.729 & 0.241 & 2.215 & 1.168 & 2.281 & 0.247 & 1.214 \\
2.51  & $-$0.40 & 0.977 & 0.731 & 0.245 & 2.194 & 1.139 & 2.290 & 0.260 & 1.212 \\
2.81  & $-$0.40 & 0.958 & 0.720 & 0.238 & 2.174 & 1.129 & 2.249 & 0.255 & 1.209 \\
3.16  & $-$0.40 & 0.928 & 0.704 & 0.223 & 2.119 & 1.091 & 2.189 & 0.265 & 1.204 \\
3.54  & $-$0.40 & 0.918 & 0.698 & 0.219 & 2.105 & 1.078 & 2.169 & 0.270 & 1.203 \\
3.98  & $-$0.40 & 0.911 & 0.694 & 0.216 & 2.110 & 1.074 & 2.162 & 0.265 & 1.202 \\
4.46  & $-$0.40 & 0.906 & 0.693 & 0.213 & 2.091 & 1.054 & 2.148 & 0.270 & 1.200 \\
5.01  & $-$0.40 & 0.908 & 0.693 & 0.215 & 2.100 & 1.044 & 2.149 & 0.275 & 1.199 \\
5.62  & $-$0.40 & 0.904 & 0.690 & 0.213 & 2.097 & 1.034 & 2.142 & 0.275 & 1.197 \\
6.31  & $-$0.40 & 0.894 & 0.683 & 0.211 & 2.088 & 1.004 & 2.127 & 0.277 & 1.193 \\
7.08  & $-$0.40 & 0.896 & 0.684 & 0.212 & 2.088 & 0.993 & 2.127 & 0.279 & 1.192 \\
7.94  & $-$0.40 & 0.894 & 0.682 & 0.211 & 2.091 & 0.983 & 2.119 & 0.282 & 1.190 \\
8.91  & $-$0.40 & 0.890 & 0.680 & 0.210 & 2.083 & 0.968 & 2.112 & 0.287 & 1.189 \\
10.00 & $-$0.40 & 0.892 & 0.681 & 0.211 & 2.093 & 0.958 & 2.114 & 0.286 & 1.188 \\
11.22 & $-$0.40 & 0.894 & 0.681 & 0.212 & 2.099 & 0.948 & 2.120 & 0.293 & 1.187 \\
12.59 & $-$0.40 & 0.887 & 0.678 & 0.209 & 2.110 & 0.937 & 2.120 & 0.292 & 1.185 \\
14.13 & $-$0.40 & 0.885 & 0.677 & 0.208 & 2.109 & 0.922 & 2.117 & 0.294 & 1.184 \\
 & & & & & & & & & \\
1.00  & 0.00	& 1.120 & 0.783 & 0.337 & 2.207 & 1.301 & 2.158 & 0.062 & 1.185 \\
1.12  & 0.00	& 1.128 & 0.788 & 0.340 & 2.246 & 1.341 & 2.174 & 0.057 & 1.185 \\
1.25  & 0.00	& 1.140 & 0.795 & 0.345 & 2.248 & 1.319 & 2.205 & 0.046 & 1.186 \\
1.41  & 0.00	& 1.148 & 0.802 & 0.346 & 2.239 & 1.289 & 2.221 & 0.064 & 1.187 \\
1.58  & 0.00	& 1.107 & 0.803 & 0.304 & 2.224 & 1.122 & 2.406 & 0.327 & 1.227 \\
1.77  & 0.00	& 1.067 & 0.782 & 0.284 & 2.229 & 1.143 & 2.376 & 0.323 & 1.224 \\
1.99  & 0.00	& 1.046 & 0.768 & 0.278 & 2.197 & 1.112 & 2.349 & 0.338 & 1.218 \\
2.23  & 0.00	& 1.036 & 0.761 & 0.275 & 2.170 & 1.077 & 2.310 & 0.352 & 1.213 \\
2.51  & 0.00	& 1.030 & 0.760 & 0.270 & 2.197 & 1.106 & 2.315 & 0.348 & 1.213 \\
2.81  & 0.00	& 1.025 & 0.755 & 0.270 & 2.149 & 1.060 & 2.264 & 0.370 & 1.211 \\
3.16  & 0.00	& 1.009 & 0.746 & 0.263 & 2.128 & 1.032 & 2.224 & 0.378 & 1.207 \\
3.54  & 0.00	& 1.024 & 0.757 & 0.267 & 2.183 & 1.063 & 2.279 & 0.368 & 1.208 \\
3.98  & 0.00	& 1.015 & 0.750 & 0.264 & 2.157 & 1.030 & 2.242 & 0.383 & 1.204 \\
4.46  & 0.00	& 1.006 & 0.744 & 0.262 & 2.142 & 1.010 & 2.222 & 0.392 & 1.202 \\
5.01  & 0.00	& 1.004 & 0.743 & 0.261 & 2.147 & 1.012 & 2.215 & 0.395 & 1.202 \\
5.62  & 0.00	& 1.005 & 0.742 & 0.263 & 2.142 & 0.990 & 2.207 & 0.405 & 1.201 \\
6.31  & 0.00	& 0.977 & 0.724 & 0.252 & 2.119 & 0.953 & 2.154 & 0.416 & 1.195 \\
7.08  & 0.00	& 0.973 & 0.722 & 0.250 & 2.138 & 0.955 & 2.163 & 0.417 & 1.195 \\
7.94  & 0.00	& 0.974 & 0.722 & 0.251 & 2.153 & 0.956 & 2.161 & 0.420 & 1.194 \\
8.91  & 0.00	& 0.973 & 0.723 & 0.250 & 2.171 & 0.947 & 2.181 & 0.416 & 1.192 \\
10.00 & 0.00	& 0.963 & 0.715 & 0.247 & 2.158 & 0.927 & 2.138 & 0.426 & 1.190 \\
11.22 & 0.00	& 0.967 & 0.718 & 0.248 & 2.192 & 0.933 & 2.178 & 0.427 & 1.190 \\
12.59 & 0.00	& 0.961 & 0.714 & 0.247 & 2.172 & 0.896 & 2.151 & 0.439 & 1.188 \\
14.13 & 0.00	& 0.963 & 0.715 & 0.247 & 2.203 & 0.902 & 2.168 & 0.435 & 1.188 \\
 & & & & & & & & & \\
1.00  & 0.20	& 1.181 & 0.812 & 0.368 & 2.191 & 1.320 & 2.117 & 0.016 & 1.176 \\
1.12  & 0.20	& 1.157 & 0.810 & 0.347 & 2.205 & 1.388 & 2.097 & 0.022 & 1.176 \\
1.25  & 0.20	& 1.124 & 0.789 & 0.335 & 2.149 & 1.315 & 2.050 & 0.059 & 1.175 \\
1.41  & 0.20	& 1.116 & 0.790 & 0.326 & 2.144 & 1.307 & 2.062 & 0.075 & 1.177 \\
1.58  & 0.20	& 1.181 & 0.850 & 0.330 & 2.269 & 1.130 & 2.486 & 0.343 & 1.229 \\
1.77  & 0.20	& 1.031 & 0.759 & 0.271 & 2.172 & 1.076 & 2.297 & 0.340 & 1.211 \\
1.99  & 0.20	& 1.018 & 0.752 & 0.266 & 2.198 & 1.091 & 2.290 & 0.339 & 1.210 \\
2.23  & 0.20	& 1.003 & 0.744 & 0.259 & 2.152 & 1.048 & 2.226 & 0.363 & 1.206 \\
2.51  & 0.20	& 0.990 & 0.736 & 0.253 & 2.095 & 1.003 & 2.167 & 0.383 & 1.203 \\
2.81  & 0.20	& 0.993 & 0.740 & 0.252 & 2.090 & 0.986 & 2.199 & 0.386 & 1.200 \\
3.16  & 0.20	& 0.986 & 0.735 & 0.251 & 2.112 & 0.999 & 2.149 & 0.394 & 1.201 \\
3.54  & 0.20	& 0.978 & 0.734 & 0.243 & 2.153 & 1.025 & 2.176 & 0.386 & 1.201 \\
3.98  & 0.20	& 0.974 & 0.731 & 0.243 & 2.113 & 0.979 & 2.150 & 0.405 & 1.198 \\
4.46  & 0.20	& 0.969 & 0.728 & 0.240 & 2.108 & 0.964 & 2.151 & 0.409 & 1.196 \\
5.01  & 0.20	& 0.959 & 0.720 & 0.239 & 2.084 & 0.930 & 2.117 & 0.421 & 1.192 \\
5.62  & 0.20	& 0.960 & 0.721 & 0.239 & 2.085 & 0.913 & 2.128 & 0.430 & 1.191 \\
6.31  & 0.20	& 0.958 & 0.721 & 0.236 & 2.129 & 0.927 & 2.173 & 0.423 & 1.191 \\
7.08  & 0.20	& 0.954 & 0.717 & 0.237 & 2.118 & 0.891 & 2.143 & 0.443 & 1.188 \\
7.94  & 0.20	& 0.953 & 0.718 & 0.235 & 2.248 & 0.975 & 2.206 & 0.415 & 1.192 \\
8.91  & 0.20	& 0.943 & 0.712 & 0.231 & 2.224 & 0.929 & 2.184 & 0.439 & 1.188 \\
10.00 & 0.20	& 0.940 & 0.708 & 0.232 & 2.174 & 0.866 & 2.144 & 0.463 & 1.184 \\
11.22 & 0.20	& 0.938 & 0.708 & 0.229 & 2.190 & 0.855 & 2.161 & 0.470 & 1.182 \\
12.59 & 0.20	& 0.942 & 0.710 & 0.231 & 2.190 & 0.837 & 2.160 & 0.477 & 1.181 \\
14.13 & 0.20	& 0.944 & 0.713 & 0.230 & 2.244 & 0.852 & 2.236 & 0.462 & 1.181 \\
\end{longtable}
\end{footnotesize}

\begin{footnotesize}
\begin{longtable}{rrrrrrrrrr}
 \caption{Integrated colours and abundances for the GirS models as a function of age and metallicity}
 \label{GIRSparameters}	\\
 \hline
 \hline
 {\bf Age ($\mathrm{Gyr}$)} & {\bf [Z/Z$_{\odot}$]} & {\bf (J-K)} & {\bf (J-H)} & {\bf (H-K)} & {\bf Na I (\AA)} & {\bf Fe I (\AA)} & {\bf Ca I (\AA)} & {\bf Mg I (\AA)} & {\bf D$_{\mathrm{CO}}$ (mag)} \\ 
 \hline
 \hline
 \endfirsthead
 \hline
 \hline
 \endlastfoot
 \multicolumn{10}{c}{\tablename\ \thetable\ - Continued from previous page} \\
 \multicolumn{10}{c}{ } \\
 \hline
 \hline
 {\bf Age ($\mathrm{Gyr}$)} & {\bf [Z/Z$_{\odot}$]} & {\bf (J-K)} & {\bf (J-H)} & {\bf (H-K)} & {\bf Na I (\AA)} & {\bf Fe I (\AA)} & {\bf Ca I (\AA)} & {\bf Mg I (\AA)} & {\bf D$_{\mathrm{CO}}$ (mag)} \\ 
 \hline
 \hline
 \endhead
 \hline
 \hline
 \endfoot
1.00  & $-$0.70 & 0.906 & 0.667 & 0.238 & 2.221 & 1.229 & 2.159 & 0.203 & 1.213 \\
1.12  & $-$0.70 & 0.900 & 0.663 & 0.237 & 2.205 & 1.214 & 2.144 & 0.206 & 1.212 \\
1.26  & $-$0.70 & 0.891 & 0.661 & 0.229 & 2.140 & 1.171 & 2.098 & 0.217 & 1.207 \\
1.41  & $-$0.70 & 0.875 & 0.658 & 0.217 & 2.042 & 1.102 & 2.028 & 0.239 & 1.201 \\
1.58  & $-$0.70 & 0.922 & 0.689 & 0.232 & 2.187 & 1.178 & 2.177 & 0.241 & 1.216 \\
1.78  & $-$0.70 & 0.898 & 0.673 & 0.224 & 2.119 & 1.122 & 2.116 & 0.249 & 1.209 \\
2.00  & $-$0.70 & 0.883 & 0.665 & 0.218 & 2.061 & 1.075 & 2.065 & 0.259 & 1.204 \\
2.24  & $-$0.70 & 0.877 & 0.662 & 0.214 & 2.035 & 1.052 & 2.044 & 0.263 & 1.201 \\
2.51  & $-$0.70 & 0.869 & 0.657 & 0.212 & 2.016 & 1.036 & 2.024 & 0.266 & 1.199 \\
2.82  & $-$0.70 & 0.861 & 0.651 & 0.209 & 1.986 & 1.010 & 1.996 & 0.271 & 1.197 \\
3.16  & $-$0.70 & 0.858 & 0.649 & 0.208 & 1.980 & 1.000 & 1.987 & 0.272 & 1.195 \\
3.55  & $-$0.70 & 0.853 & 0.646 & 0.207 & 1.969 & 0.984 & 1.976 & 0.274 & 1.194 \\
3.98  & $-$0.70 & 0.851 & 0.644 & 0.207 & 1.961 & 0.972 & 1.967 & 0.277 & 1.193 \\
4.47  & $-$0.70 & 0.839 & 0.635 & 0.203 & 1.934 & 0.942 & 1.938 & 0.279 & 1.189 \\
5.01  & $-$0.70 & 0.839 & 0.636 & 0.203 & 1.933 & 0.933 & 1.935 & 0.282 & 1.188 \\
5.62  & $-$0.70 & 0.830 & 0.629 & 0.201 & 1.916 & 0.907 & 1.911 & 0.281 & 1.185 \\
6.31  & $-$0.70 & 0.830 & 0.629 & 0.200 & 1.914 & 0.899 & 1.907 & 0.285 & 1.184 \\
7.08  & $-$0.70 & 0.828 & 0.628 & 0.200 & 1.905 & 0.883 & 1.898 & 0.288 & 1.182 \\
7.94  & $-$0.70 & 0.828 & 0.627 & 0.200 & 1.907 & 0.870 & 1.894 & 0.290 & 1.181 \\
8.91  & $-$0.70 & 0.826 & 0.625 & 0.200 & 1.897 & 0.854 & 1.880 & 0.293 & 1.179 \\
10.00 & $-$0.70 & 0.823 & 0.623 & 0.200 & 1.897 & 0.840 & 1.873 & 0.294 & 1.177 \\
11.20 & $-$0.70 & 0.821 & 0.622 & 0.198 & 1.898 & 0.832 & 1.876 & 0.294 & 1.175 \\
12.60 & $-$0.70 & 0.818 & 0.620 & 0.198 & 1.894 & 0.811 & 1.864 & 0.298 & 1.173 \\
14.10 & $-$0.70 & 0.814 & 0.617 & 0.197 & 1.887 & 0.794 & 1.857 & 0.297 & 1.171 \\
 & & & & & & & & & \\
1.00  & $-$0.40 & 0.918 & 0.678 & 0.240 & 2.076 & 1.206 & 2.024 & 0.157 & 1.189 \\
1.12  & $-$0.40 & 0.913 & 0.677 & 0.235 & 2.054 & 1.186 & 2.012 & 0.166 & 1.188 \\
1.26  & $-$0.40 & 0.916 & 0.689 & 0.226 & 2.021 & 1.162 & 2.006 & 0.190 & 1.190 \\
1.41  & $-$0.40 & 0.956 & 0.716 & 0.240 & 2.116 & 1.241 & 2.084 & 0.174 & 1.197 \\
1.58  & $-$0.40 & 0.935 & 0.705 & 0.229 & 2.109 & 1.098 & 2.187 & 0.262 & 1.206 \\
1.78  & $-$0.40 & 0.933 & 0.705 & 0.227 & 2.111 & 1.097 & 2.192 & 0.264 & 1.206 \\
2.00  & $-$0.40 & 0.914 & 0.694 & 0.220 & 2.070 & 1.065 & 2.154 & 0.269 & 1.203 \\
2.24  & $-$0.40 & 0.906 & 0.690 & 0.215 & 2.055 & 1.053 & 2.138 & 0.271 & 1.201 \\
2.51  & $-$0.40 & 0.903 & 0.690 & 0.213 & 2.045 & 1.041 & 2.127 & 0.276 & 1.199 \\
2.82  & $-$0.40 & 0.897 & 0.687 & 0.210 & 2.048 & 1.041 & 2.119 & 0.276 & 1.199 \\
3.16  & $-$0.40 & 0.883 & 0.678 & 0.204 & 2.012 & 1.005 & 2.083 & 0.282 & 1.194 \\
3.55  & $-$0.40 & 0.877 & 0.676 & 0.201 & 2.000 & 0.990 & 2.066 & 0.286 & 1.193 \\
3.98  & $-$0.40 & 0.877 & 0.676 & 0.201 & 2.003 & 0.986 & 2.063 & 0.285 & 1.192 \\
4.47  & $-$0.40 & 0.876 & 0.675 & 0.201 & 2.010 & 0.983 & 2.068 & 0.286 & 1.191 \\
5.01  & $-$0.40 & 0.871 & 0.671 & 0.199 & 2.006 & 0.965 & 2.053 & 0.290 & 1.189 \\
5.62  & $-$0.40 & 0.872 & 0.673 & 0.199 & 2.013 & 0.962 & 2.056 & 0.291 & 1.189 \\
6.31  & $-$0.40 & 0.870 & 0.671 & 0.199 & 2.016 & 0.950 & 2.052 & 0.293 & 1.187 \\
7.08  & $-$0.40 & 0.867 & 0.668 & 0.199 & 2.027 & 0.931 & 2.055 & 0.293 & 1.184 \\
7.94  & $-$0.40 & 0.867 & 0.667 & 0.200 & 2.035 & 0.920 & 2.056 & 0.293 & 1.183 \\
8.91  & $-$0.40 & 0.867 & 0.667 & 0.199 & 2.040 & 0.914 & 2.058 & 0.298 & 1.182 \\
10.00 & $-$0.40 & 0.869 & 0.668 & 0.201 & 2.054 & 0.906 & 2.070 & 0.300 & 1.182 \\
11.20 & $-$0.40 & 0.871 & 0.669 & 0.202 & 2.066 & 0.905 & 2.080 & 0.301 & 1.181 \\
12.60 & $-$0.40 & 0.875 & 0.670 & 0.204 & 2.089 & 0.902 & 2.097 & 0.302 & 1.181 \\
14.10 & $-$0.40 & 0.879 & 0.672 & 0.206 & 2.110 & 0.898 & 2.114 & 0.303 & 1.180 \\
 & & & & & & & & & \\
1.00  & 0.00	& 0.945 & 0.687 & 0.258 & 1.955 & 1.124 & 1.898 & 0.179 & 1.169 \\
1.12  & 0.00	& 0.939 & 0.684 & 0.254 & 1.943 & 1.110 & 1.889 & 0.188 & 1.168 \\
1.26  & 0.00	& 0.937 & 0.690 & 0.246 & 1.929 & 1.095 & 1.895 & 0.206 & 1.171 \\
1.41  & 0.00	& 0.934 & 0.696 & 0.238 & 1.921 & 1.083 & 1.910 & 0.228 & 1.174 \\
1.58  & 0.00	& 0.956 & 0.715 & 0.241 & 2.035 & 1.015 & 2.154 & 0.349 & 1.201 \\
1.78  & 0.00	& 0.953 & 0.713 & 0.239 & 2.068 & 1.031 & 2.168 & 0.351 & 1.200 \\
2.00  & 0.00	& 0.931 & 0.700 & 0.230 & 2.038 & 1.011 & 2.127 & 0.362 & 1.197 \\
2.24  & 0.00	& 0.930 & 0.701 & 0.229 & 2.049 & 1.006 & 2.138 & 0.365 & 1.196 \\
2.51  & 0.00	& 0.930 & 0.701 & 0.229 & 2.042 & 0.994 & 2.127 & 0.374 & 1.194 \\
2.82  & 0.00	& 0.933 & 0.702 & 0.230 & 2.030 & 0.979 & 2.102 & 0.386 & 1.194 \\
3.16  & 0.00	& 0.936 & 0.705 & 0.230 & 2.057 & 0.985 & 2.122 & 0.391 & 1.194 \\
3.55  & 0.00	& 0.944 & 0.709 & 0.234 & 2.064 & 0.967 & 2.124 & 0.397 & 1.193 \\
3.98  & 0.00	& 0.937 & 0.706 & 0.231 & 2.054 & 0.953 & 2.104 & 0.406 & 1.191 \\
4.47  & 0.00	& 0.933 & 0.703 & 0.229 & 2.073 & 0.961 & 2.121 & 0.402 & 1.191 \\
5.01  & 0.00	& 0.936 & 0.703 & 0.232 & 2.050 & 0.930 & 2.082 & 0.418 & 1.189 \\
5.62  & 0.00	& 0.937 & 0.704 & 0.232 & 2.069 & 0.934 & 2.103 & 0.415 & 1.189 \\
6.31  & 0.00	& 0.929 & 0.697 & 0.232 & 2.073 & 0.908 & 2.079 & 0.428 & 1.186 \\
7.08  & 0.00	& 0.931 & 0.698 & 0.233 & 2.090 & 0.907 & 2.091 & 0.429 & 1.186 \\
7.94  & 0.00	& 0.934 & 0.699 & 0.234 & 2.115 & 0.909 & 2.113 & 0.427 & 1.185 \\
8.91  & 0.00	& 0.935 & 0.699 & 0.236 & 2.105 & 0.883 & 2.096 & 0.436 & 1.185 \\
10.00 & 0.00	& 0.933 & 0.697 & 0.236 & 2.105 & 0.868 & 2.081 & 0.445 & 1.183 \\
11.20 & 0.00	& 0.943 & 0.704 & 0.239 & 2.162 & 0.893 & 2.148 & 0.436 & 1.184 \\
12.60 & 0.00	& 0.946 & 0.701 & 0.245 & 2.109 & 0.828 & 2.059 & 0.472 & 1.182 \\
14.10 & 0.00	& 0.954 & 0.705 & 0.249 & 2.126 & 0.823 & 2.055 & 0.478 & 1.182 \\
 & & & & & & & & & \\
1.00  & 0.20	& 0.898 & 0.658 & 0.240 & 1.848 & 1.011 & 1.831 & 0.208 & 1.160 \\
1.12  & 0.20	& 0.898 & 0.658 & 0.239 & 1.837 & 0.996 & 1.827 & 0.213 & 1.160 \\
1.26  & 0.20	& 0.898 & 0.661 & 0.237 & 1.810 & 0.970 & 1.810 & 0.226 & 1.159 \\
1.41  & 0.20	& 0.905 & 0.675 & 0.229 & 1.828 & 0.979 & 1.852 & 0.244 & 1.165 \\
1.58  & 0.20	& 0.989 & 0.735 & 0.253 & 1.975 & 0.952 & 2.125 & 0.390 & 1.203 \\
1.78  & 0.20	& 0.909 & 0.687 & 0.222 & 1.891 & 0.876 & 2.024 & 0.396 & 1.188 \\
2.00  & 0.20	& 0.905 & 0.685 & 0.220 & 1.891 & 0.873 & 2.013 & 0.399 & 1.187 \\
2.24  & 0.20	& 0.902 & 0.684 & 0.218 & 1.863 & 0.838 & 1.956 & 0.427 & 1.184 \\
2.51  & 0.20	& 0.910 & 0.688 & 0.221 & 1.888 & 0.850 & 1.965 & 0.430 & 1.185 \\
2.82  & 0.20	& 0.908 & 0.691 & 0.216 & 1.927 & 0.872 & 2.005 & 0.421 & 1.186 \\
3.16  & 0.20	& 0.918 & 0.694 & 0.224 & 1.911 & 0.849 & 1.957 & 0.444 & 1.186 \\
3.55  & 0.20	& 0.923 & 0.700 & 0.223 & 1.934 & 0.861 & 1.983 & 0.443 & 1.186 \\
3.98  & 0.20	& 0.924 & 0.700 & 0.224 & 1.957 & 0.859 & 1.996 & 0.447 & 1.187 \\
4.47  & 0.20	& 0.927 & 0.703 & 0.223 & 1.974 & 0.864 & 2.017 & 0.443 & 1.186 \\
5.01  & 0.20	& 0.920 & 0.697 & 0.222 & 1.987 & 0.853 & 2.014 & 0.449 & 1.185 \\
5.62  & 0.20	& 0.925 & 0.700 & 0.224 & 2.013 & 0.857 & 2.036 & 0.450 & 1.185 \\
6.31  & 0.20	& 0.929 & 0.704 & 0.224 & 2.064 & 0.876 & 2.092 & 0.442 & 1.186 \\
7.08  & 0.20	& 0.929 & 0.702 & 0.226 & 2.066 & 0.849 & 2.082 & 0.458 & 1.184 \\
7.94  & 0.20	& 0.935 & 0.705 & 0.229 & 2.109 & 0.856 & 2.115 & 0.461 & 1.184 \\
8.91  & 0.20	& 0.935 & 0.707 & 0.228 & 2.146 & 0.861 & 2.160 & 0.457 & 1.183 \\
10.00 & 0.20	& 0.935 & 0.704 & 0.230 & 2.140 & 0.828 & 2.132 & 0.479 & 1.181 \\
11.20 & 0.20	& 0.939 & 0.708 & 0.231 & 2.167 & 0.833 & 2.158 & 0.475 & 1.181 \\
12.60 & 0.20	& 0.942 & 0.710 & 0.232 & 2.188 & 0.827 & 2.168 & 0.482 & 1.180 \\
14.10 & 0.20	& 0.947 & 0.715 & 0.232 & 2.260 & 0.861 & 2.251 & 0.466 & 1.182 \\
\end{longtable}
\end{footnotesize}

\begin{footnotesize}
\begin{longtable}{rrrrrrrrrr}
 \caption{Integrated colours and abundances for the BaSS models as a function of age and metallicity}
 \label{BASSparameters}	\\
 \hline
 \hline
 {\bf Age ($\mathrm{Gyr}$)} & {\bf [Z/Z$_{\odot}$]} & {\bf (J-K)} & {\bf (J-H)} & {\bf (H-K)} & {\bf Na I (\AA)} & {\bf Fe I (\AA)} & {\bf Ca I (\AA)} & {\bf Mg I (\AA)} & {\bf D$_{\mathrm{CO}}$ (mag)} \\ 
 \hline
 \hline
 \endfirsthead
 \hline
 \hline
 \endlastfoot
 \multicolumn{10}{c}{\tablename\ \thetable\ - Continued from previous page} \\
 \multicolumn{10}{c}{ } \\
 \hline
 \hline
 {\bf Age ($\mathrm{Gyr}$)} & {\bf [Z/Z$_{\odot}$]} & {\bf (J-K)} & {\bf (J-H)} & {\bf (H-K)} & {\bf Na I (\AA)} & {\bf Fe I (\AA)} & {\bf Ca I (\AA)} & {\bf Mg I (\AA)} & {\bf D$_{\mathrm{CO}}$ (mag)} \\ 
 \hline
 \hline
 \endhead
 \hline
 \hline
 \endfoot
1.00  & $-$0.70 & 0.867 & 0.659 & 0.208 & 1.891 & 1.067 & 1.907 & 0.216 & 1.186 \\  
1.25  & $-$0.70 & 0.849 & 0.649 & 0.199 & 1.841 & 1.023 & 1.874 & 0.233 & 1.184 \\  
1.50  & $-$0.70 & 0.842 & 0.646 & 0.196 & 1.825 & 1.004 & 1.861 & 0.243 & 1.183 \\  
1.75  & $-$0.70 & 0.824 & 0.640 & 0.183 & 1.834 & 0.957 & 1.899 & 0.287 & 1.191 \\  
2.00  & $-$0.70 & 0.821 & 0.637 & 0.183 & 1.825 & 0.948 & 1.882 & 0.293 & 1.190 \\  
2.25  & $-$0.70 & 0.819 & 0.636 & 0.183 & 1.812 & 0.938 & 1.869 & 0.297 & 1.190 \\  
2.50  & $-$0.70 & 0.814 & 0.632 & 0.181 & 1.798 & 0.927 & 1.857 & 0.299 & 1.188 \\  
2.75  & $-$0.70 & 0.813 & 0.631 & 0.181 & 1.799 & 0.927 & 1.855 & 0.300 & 1.189 \\  
3.00  & $-$0.70 & 0.810 & 0.629 & 0.180 & 1.797 & 0.925 & 1.855 & 0.300 & 1.188 \\  
3.25  & $-$0.70 & 0.807 & 0.628 & 0.178 & 1.789 & 0.920 & 1.853 & 0.300 & 1.188 \\  
3.50  & $-$0.70 & 0.807 & 0.628 & 0.178 & 1.792 & 0.922 & 1.856 & 0.302 & 1.188 \\  
3.75  & $-$0.70 & 0.808 & 0.629 & 0.179 & 1.792 & 0.921 & 1.857 & 0.303 & 1.188 \\  
4.00  & $-$0.70 & 0.806 & 0.627 & 0.178 & 1.788 & 0.917 & 1.852 & 0.305 & 1.188 \\  
4.50  & $-$0.70 & 0.810 & 0.630 & 0.180 & 1.790 & 0.919 & 1.856 & 0.308 & 1.189 \\  
5.00  & $-$0.70 & 0.812 & 0.631 & 0.181 & 1.794 & 0.921 & 1.858 & 0.309 & 1.189 \\  
5.50  & $-$0.70 & 0.809 & 0.628 & 0.181 & 1.787 & 0.914 & 1.847 & 0.311 & 1.188 \\  
6.00  & $-$0.70 & 0.813 & 0.631 & 0.182 & 1.795 & 0.918 & 1.855 & 0.312 & 1.188 \\  
6.50  & $-$0.70 & 0.816 & 0.632 & 0.183 & 1.800 & 0.920 & 1.859 & 0.313 & 1.189 \\  
7.00  & $-$0.70 & 0.820 & 0.635 & 0.185 & 1.806 & 0.922 & 1.863 & 0.315 & 1.189 \\  
7.50  & $-$0.70 & 0.822 & 0.635 & 0.186 & 1.807 & 0.921 & 1.863 & 0.317 & 1.189 \\  
8.00  & $-$0.70 & 0.824 & 0.637 & 0.187 & 1.811 & 0.922 & 1.866 & 0.317 & 1.189 \\  
8.50  & $-$0.70 & 0.828 & 0.639 & 0.189 & 1.818 & 0.926 & 1.871 & 0.317 & 1.189 \\  
9.00  & $-$0.70 & 0.831 & 0.641 & 0.190 & 1.825 & 0.929 & 1.877 & 0.316 & 1.189 \\  
9.50  & $-$0.70 & 0.834 & 0.642 & 0.192 & 1.831 & 0.931 & 1.880 & 0.318 & 1.190 \\  
10.00 & $-$0.70 & 0.836 & 0.642 & 0.193 & 1.836 & 0.933 & 1.883 & 0.319 & 1.190 \\  
11.00 & $-$0.70 & 0.840 & 0.645 & 0.195 & 1.849 & 0.940 & 1.894 & 0.319 & 1.191 \\  
12.00 & $-$0.70 & 0.840 & 0.644 & 0.196 & 1.857 & 0.942 & 1.898 & 0.319 & 1.191 \\  
13.00 & $-$0.70 & 0.839 & 0.642 & 0.197 & 1.861 & 0.943 & 1.898 & 0.319 & 1.191 \\  
14.00 & $-$0.70 & 0.838 & 0.641 & 0.197 & 1.871 & 0.949 & 1.904 & 0.317 & 1.192 \\ 
 & & & & & & & & & \\
1.00  & $-$0.40 & 0.920 & 0.691 & 0.228 & 1.942 & 1.172 & 1.941 & 0.158 & 1.181 \\  
1.25  & $-$0.40 & 0.898 & 0.682 & 0.215 & 1.912 & 1.130 & 1.938 & 0.188 & 1.183 \\  
1.50  & $-$0.40 & 0.883 & 0.676 & 0.206 & 1.883 & 1.094 & 1.927 & 0.212 & 1.183 \\  
1.75  & $-$0.40 & 0.860 & 0.669 & 0.191 & 1.882 & 0.992 & 2.005 & 0.294 & 1.194 \\  
2.00  & $-$0.40 & 0.857 & 0.667 & 0.189 & 1.874 & 0.986 & 1.997 & 0.298 & 1.193 \\  
2.25  & $-$0.40 & 0.856 & 0.667 & 0.189 & 1.876 & 0.986 & 1.997 & 0.298 & 1.193 \\  
2.50  & $-$0.40 & 0.856 & 0.667 & 0.189 & 1.873 & 0.983 & 1.994 & 0.299 & 1.193 \\  
2.75  & $-$0.40 & 0.856 & 0.666 & 0.189 & 1.872 & 0.980 & 1.993 & 0.299 & 1.193 \\  
3.00  & $-$0.40 & 0.855 & 0.666 & 0.189 & 1.871 & 0.979 & 1.991 & 0.300 & 1.192 \\  
3.25  & $-$0.40 & 0.857 & 0.667 & 0.189 & 1.878 & 0.982 & 1.998 & 0.300 & 1.193 \\  
3.50  & $-$0.40 & 0.859 & 0.668 & 0.190 & 1.887 & 0.988 & 2.004 & 0.299 & 1.193 \\  
3.75  & $-$0.40 & 0.858 & 0.667 & 0.190 & 1.884 & 0.984 & 2.000 & 0.301 & 1.193 \\  
4.00  & $-$0.40 & 0.860 & 0.669 & 0.191 & 1.891 & 0.988 & 2.007 & 0.300 & 1.193 \\  
4.50  & $-$0.40 & 0.863 & 0.671 & 0.192 & 1.899 & 0.990 & 2.015 & 0.301 & 1.193 \\  
5.00  & $-$0.40 & 0.864 & 0.671 & 0.193 & 1.905 & 0.991 & 2.020 & 0.301 & 1.193 \\  
5.50  & $-$0.40 & 0.868 & 0.674 & 0.194 & 1.916 & 0.998 & 2.029 & 0.301 & 1.193 \\  
6.00  & $-$0.40 & 0.872 & 0.676 & 0.195 & 1.925 & 1.001 & 2.037 & 0.302 & 1.194 \\  
6.50  & $-$0.40 & 0.873 & 0.677 & 0.196 & 1.928 & 1.000 & 2.040 & 0.303 & 1.193 \\  
7.00  & $-$0.40 & 0.875 & 0.677 & 0.197 & 1.934 & 1.002 & 2.047 & 0.304 & 1.194 \\  
7.50  & $-$0.40 & 0.877 & 0.679 & 0.198 & 1.940 & 1.005 & 2.053 & 0.303 & 1.194 \\  
8.00  & $-$0.40 & 0.881 & 0.682 & 0.199 & 1.948 & 1.008 & 2.060 & 0.303 & 1.194 \\  
8.50  & $-$0.40 & 0.885 & 0.684 & 0.200 & 1.956 & 1.011 & 2.069 & 0.303 & 1.194 \\  
9.00  & $-$0.40 & 0.887 & 0.685 & 0.202 & 1.964 & 1.014 & 2.076 & 0.303 & 1.194 \\  
9.50  & $-$0.40 & 0.889 & 0.686 & 0.203 & 1.969 & 1.015 & 2.081 & 0.303 & 1.194 \\  
10.00 & $-$0.40 & 0.892 & 0.687 & 0.204 & 1.976 & 1.017 & 2.088 & 0.303 & 1.195 \\  
11.00 & $-$0.40 & 0.895 & 0.689 & 0.206 & 1.994 & 1.026 & 2.102 & 0.301 & 1.195 \\  
12.00 & $-$0.40 & 0.894 & 0.688 & 0.205 & 1.993 & 1.020 & 2.102 & 0.306 & 1.195 \\  
13.00 & $-$0.40 & 0.895 & 0.689 & 0.206 & 2.009 & 1.030 & 2.116 & 0.303 & 1.195 \\  
14.00 & $-$0.40 & 0.897 & 0.690 & 0.207 & 2.015 & 1.030 & 2.124 & 0.305 & 1.195 \\
 & & & & & & & & & \\
1.00  & 0.00	& 0.994 & 0.724 & 0.269 & 2.016 & 1.166 & 2.020 & 0.178 & 1.184 \\  
1.25  & 0.00	& 0.961 & 0.708 & 0.252 & 1.984 & 1.103 & 2.039 & 0.222 & 1.186 \\  
1.50  & 0.00	& 0.939 & 0.701 & 0.238 & 1.942 & 1.083 & 1.984 & 0.264 & 1.186 \\  
1.75  & 0.00	& 0.920 & 0.695 & 0.224 & 1.923 & 0.984 & 2.029 & 0.375 & 1.198 \\  
2.00  & 0.00	& 0.920 & 0.695 & 0.224 & 1.913 & 0.971 & 2.020 & 0.383 & 1.197 \\  
2.25  & 0.00	& 0.921 & 0.696 & 0.225 & 1.910 & 0.965 & 2.019 & 0.387 & 1.197 \\  
2.50  & 0.00	& 0.925 & 0.697 & 0.227 & 1.915 & 0.966 & 2.016 & 0.392 & 1.196 \\  
2.75  & 0.00	& 0.927 & 0.699 & 0.228 & 1.919 & 0.965 & 2.026 & 0.394 & 1.196 \\  
3.00  & 0.00	& 0.928 & 0.699 & 0.228 & 1.918 & 0.964 & 2.021 & 0.398 & 1.196 \\  
3.25  & 0.00	& 0.932 & 0.703 & 0.229 & 1.945 & 0.984 & 2.054 & 0.390 & 1.197 \\  
3.50  & 0.00	& 0.933 & 0.704 & 0.229 & 1.956 & 0.990 & 2.066 & 0.390 & 1.197 \\  
3.75  & 0.00	& 0.929 & 0.701 & 0.228 & 1.949 & 0.982 & 2.054 & 0.396 & 1.196 \\  
4.00  & 0.00	& 0.930 & 0.700 & 0.230 & 1.933 & 0.965 & 2.025 & 0.408 & 1.195 \\  
4.50  & 0.00	& 0.931 & 0.700 & 0.230 & 1.933 & 0.962 & 2.020 & 0.415 & 1.196 \\  
5.00  & 0.00	& 0.933 & 0.700 & 0.233 & 1.927 & 0.951 & 2.002 & 0.425 & 1.195 \\  
5.50  & 0.00	& 0.938 & 0.702 & 0.236 & 1.919 & 0.936 & 1.989 & 0.436 & 1.195 \\  
6.00  & 0.00	& 0.942 & 0.702 & 0.240 & 1.900 & 0.917 & 1.944 & 0.454 & 1.194 \\  
6.50  & 0.00	& 0.942 & 0.704 & 0.237 & 1.934 & 0.940 & 2.003 & 0.439 & 1.195 \\  
7.00  & 0.00	& 0.944 & 0.706 & 0.238 & 1.938 & 0.944 & 1.994 & 0.442 & 1.195 \\  
7.50  & 0.00	& 0.946 & 0.706 & 0.239 & 1.932 & 0.936 & 1.982 & 0.448 & 1.195 \\  
8.00  & 0.00	& 0.948 & 0.707 & 0.241 & 1.931 & 0.929 & 1.981 & 0.452 & 1.194 \\  
8.50  & 0.00	& 0.949 & 0.705 & 0.243 & 1.905 & 0.903 & 1.939 & 0.470 & 1.193 \\  
9.00  & 0.00	& 0.947 & 0.701 & 0.245 & 1.872 & 0.876 & 1.882 & 0.488 & 1.192 \\  
9.50  & 0.00	& 0.946 & 0.707 & 0.238 & 1.955 & 0.942 & 2.008 & 0.448 & 1.195 \\  
10.00 & 0.00	& 0.947 & 0.707 & 0.239 & 1.955 & 0.940 & 2.010 & 0.450 & 1.195 \\  
11.00 & 0.00	& 0.951 & 0.710 & 0.241 & 1.966 & 0.941 & 2.030 & 0.450 & 1.195 \\  
12.00 & 0.00	& 0.951 & 0.711 & 0.240 & 1.983 & 0.951 & 2.055 & 0.446 & 1.195 \\  
13.00 & 0.00	& 0.958 & 0.717 & 0.240 & 2.013 & 0.977 & 2.090 & 0.434 & 1.197 \\  
14.00 & 0.00	& 0.958 & 0.719 & 0.239 & 2.021 & 0.979 & 2.101 & 0.434 & 1.197 \\
 & & & & & & & & & \\
1.00  & 0.20	& 0.987 & 0.718 & 0.268 & 1.984 & 1.085 & 2.064 & 0.175 & 1.178 \\  
1.25  & 0.20	& 0.977 & 0.719 & 0.257 & 1.956 & 1.066 & 2.040 & 0.220 & 1.183 \\  
1.50  & 0.20	& 0.960 & 0.710 & 0.250 & 1.890 & 1.000 & 1.949 & 0.286 & 1.183 \\  
1.75  & 0.20	& 0.925 & 0.699 & 0.226 & 1.890 & 0.950 & 1.983 & 0.392 & 1.196 \\  
2.00  & 0.20	& 0.931 & 0.701 & 0.229 & 1.883 & 0.938 & 1.967 & 0.404 & 1.195 \\  
2.25  & 0.20	& 0.933 & 0.702 & 0.230 & 1.878 & 0.930 & 1.959 & 0.413 & 1.195 \\  
2.50  & 0.20	& 0.933 & 0.703 & 0.229 & 1.881 & 0.930 & 1.974 & 0.414 & 1.194 \\  
2.75  & 0.20	& 0.935 & 0.703 & 0.232 & 1.868 & 0.914 & 1.947 & 0.427 & 1.194 \\  
3.00  & 0.20	& 0.937 & 0.705 & 0.231 & 1.879 & 0.920 & 1.971 & 0.425 & 1.194 \\  
3.25  & 0.20	& 0.936 & 0.705 & 0.230 & 1.881 & 0.919 & 1.977 & 0.426 & 1.194 \\  
3.50  & 0.20	& 0.939 & 0.705 & 0.234 & 1.860 & 0.897 & 1.944 & 0.442 & 1.194 \\  
3.75  & 0.20	& 0.941 & 0.706 & 0.234 & 1.881 & 0.911 & 1.970 & 0.436 & 1.194 \\  
4.00  & 0.20	& 0.940 & 0.705 & 0.235 & 1.874 & 0.900 & 1.962 & 0.443 & 1.194 \\  
4.50  & 0.20	& 0.941 & 0.704 & 0.237 & 1.847 & 0.871 & 1.923 & 0.462 & 1.192 \\  
5.00  & 0.20	& 0.943 & 0.704 & 0.239 & 1.847 & 0.865 & 1.912 & 0.470 & 1.192 \\  
5.50  & 0.20	& 0.941 & 0.703 & 0.238 & 1.851 & 0.861 & 1.921 & 0.472 & 1.192 \\  
6.00  & 0.20	& 0.944 & 0.705 & 0.238 & 1.859 & 0.864 & 1.928 & 0.473 & 1.192 \\  
6.50  & 0.20	& 0.946 & 0.706 & 0.239 & 1.870 & 0.869 & 1.945 & 0.471 & 1.192 \\  
7.00  & 0.20	& 0.946 & 0.705 & 0.241 & 1.850 & 0.846 & 1.911 & 0.488 & 1.191 \\  
7.50  & 0.20	& 0.949 & 0.707 & 0.241 & 1.867 & 0.855 & 1.934 & 0.486 & 1.192 \\  
8.00  & 0.20	& 0.949 & 0.708 & 0.240 & 1.880 & 0.864 & 1.946 & 0.484 & 1.192 \\  
8.50  & 0.20	& 0.949 & 0.709 & 0.240 & 1.890 & 0.870 & 1.955 & 0.483 & 1.193 \\  
9.00  & 0.20	& 0.952 & 0.712 & 0.240 & 1.904 & 0.875 & 1.977 & 0.482 & 1.193 \\  
9.50  & 0.20	& 0.953 & 0.712 & 0.241 & 1.912 & 0.874 & 1.986 & 0.485 & 1.192 \\  
10.00 & 0.20	& 0.956 & 0.714 & 0.242 & 1.916 & 0.871 & 1.994 & 0.486 & 1.193 \\  
11.00 & 0.20	& 0.959 & 0.716 & 0.243 & 1.928 & 0.875 & 2.000 & 0.490 & 1.193 \\  
12.00 & 0.20	& 0.965 & 0.719 & 0.245 & 1.941 & 0.880 & 2.012 & 0.492 & 1.193 \\  
13.00 & 0.20	& 0.968 & 0.721 & 0.247 & 1.946 & 0.876 & 2.015 & 0.498 & 1.193 \\  
14.00 & 0.20	& 0.975 & 0.725 & 0.250 & 1.953 & 0.874 & 2.025 & 0.501 & 1.193 \\ 
\end{longtable}
\end{footnotesize}


\label{lastpage}

\end{document}